\documentclass[11pt,prd,onecolumn,amsmath,amssymb,aps,floats,floatfix]{revtex4-2}
\usepackage[colorlinks=true,urlcolor=blue,anchorcolor=blue,citecolor=blue,filecolor=blue,linkcolor=blue,menucolor=blue,linktocpage=true]{hyperref} 

\usepackage[inline]{enumitem}
\usepackage[multidot]{grffile}  
\usepackage{dcolumn}
\usepackage{bm}
\usepackage{amsmath}
\usepackage{amsfonts}
\usepackage{amssymb}
\usepackage{color}
\usepackage[table]{xcolor}
\usepackage{float}
\usepackage{latexsym}
\usepackage{slashed} 
\usepackage{pstricks}
\usepackage{indentfirst}
\usepackage{mathrsfs}
\usepackage{multirow}
\usepackage{epsfig,psfrag}
\usepackage{graphicx}
\usepackage{enumitem}
\usepackage{geometry,amssymb,yfonts}
\usepackage{yhmath}
\usepackage{anysize}
\usepackage{subfigure}
\usepackage{mathtools}
\usepackage{setspace} 
\usepackage[utf8]{inputenc} 
\usepackage[scientific-notation=true]{siunitx} 
\usepackage{feynmp}

\graphicspath{{fig/}}

\allowdisplaybreaks 

\newcommand{\Uone}{\mathrm{U}(1)}
\newcommand{\UoneX}{\mathrm{U}(1)_\mathrm{X}}
\newcommand{\UoneY}{\mathrm{U}(1)_\mathrm{Y}}

\newcommand{\SUtwoL}{\mathrm{SU}(2)_\mathrm{L}}

\begin{document}

\title{Complex scalar dark matter in a new gauged $\Uone$ symmetry with kinetic and direct mixings }

\author{Yu-Hang Su$^{a}$}
\author{Chengfeng Cai$^{a,b}$}
\author{Yu-Pan Zeng$^{c,a}$}\email[Corresponding author. ]{zengyupan@gdou.edu.cn}
\author{Hong-Hao Zhang$^{a}$}\email[Corresponding author. ]{zhh98@mail.sysu.edu.cn}

\affiliation{$^a$School of Physics, Sun Yat-sen University, Guangzhou 510275, China}
\affiliation{$^b$School of Science, Sun Yat-Sen University, Shenzhen 518107, China}
\affiliation{$^c$School of Electronics and Information Engineering, Guangdong Ocean University, Zhanjiang 524088, China}

\begin{abstract}
We propose a scalar dark matter model featuring a hidden gauge symmetry, denoted as $\UoneX$, with two complex scalars, $\Phi$ and $S$. In this framework, $\Phi$ spontaneously breaks the $\UoneX$ gauge symmetry, while $S$ serves as a viable dark matter candidate. Particularly, the kinetic and direct mixings between the $\UoneX$ and $\UoneY$ gauge groups provide a portal between dark matter and the Standard Model particles. These mixings offer a plausible explanation for the W boson mass anomaly observed by the CDF Collaboration. We study the comprehensive phenomenological constraints of this model from colliders and dark matter detection experiments, including $Z'$ searches at the LHC, the $125~\mathrm{GeV}$ Higgs boson measurements, the relic density of dark matter and the indirect detection of dark matter annihilation. By randomly scanning the parameter space, we find that the regions where $m_{Z'} \gtrsim 4750~\mathrm{GeV}$ and $m_{Z'} \lesssim 4750~\mathrm{GeV}$ for $g_x$ close to $1$ remain viable and can be tested by future experiments.

\end{abstract}

\maketitle
\tableofcontents

\section{Introduction}
Numerous astrophysical and cosmological observations today strongly support the existence of dark matter (DM)~\cite{Jungman:1995df, Bertone:2004pz, Feng:2010gw}, yet direct experimental confirmation remains elusive. It is well known that the couplings between DM particles and those of the standard model (SM) are comparable in magnitude to the weak force within the weakly interacting massive particle (WIMP) scenario. WIMPs are thermally produced in the early universe and subsequently decoupled from the thermal bath during the freeze-out epoch~\cite{Lee:1977ua}. 

Currently, a lot of direct detection experiments are designed to search for such DM particles, as the current and forthcoming experimental capabilities include the predicted DM mass range proposed by various theories. However, no conclusive signal has been detected so far. Consequently, many theoretical models have been ruled out due to the stringent experimental constraints on the scattering cross section between DM and nucleons~\cite{PandaX-4T:2021bab, LZ:2022ufs}. 

Apart from the DM problem, there are also particle physics anomalies awaiting resolution. In the SM, the mass of the W boson serves as a crucial and fundamental parameter. Its measurement not only provides valuable insights, but also offers a gateway to explore physics beyond the SM (BSM). In recent years, there has been increasing emphasis on precise measurements of this parameter. In 2022, the Collider Detector at Fermilab (CDF) Collaboration conducted a new measurement of the W boson mass, yielding $80.4335 \pm 0.0094~\mathrm{GeV}$~\cite{CDF:2022hxs}. This result deviates from the SM prediction by seven standard deviations, prompting the proposal of numerous theories to explain such an anomaly.  An elegant example is the $Z'$ model, which introduces a new neutral gauge boson, denoted $Z'$~\cite{Zeng:2022lkk, Cai:2022cti, Zhang:2022nnh}. This boson interacts with the $Z$ boson through a relatively weak interaction, thus evading the stringent constraints imposed by the searches $Z'$ at the LHC. It is noteworthy that in 2023, the ATLAS Collaboration reported a different result following an improved re-analysis of their measurement data. They determined the W boson mass to be $80.360 \pm 0.016~\mathrm{GeV}$, showing no deviation from the SM prediction~\cite{ATLAS:2023fsi}. Several papers have discussed the implications and potential explanations arising from these two measurements~\cite{Amoroso:2023pey, Zhu:2023qyt}. Our focus will remain on further measurements of the W boson mass, including the utilization of data from different center-of-mass energies. 

When establishing new physical models, the minimal gauge extension to the SM is achieved by introducing an extra $\Uone$ gauge symmetry, denoted $\UoneX$. A natural proposition is that this symmetry originates from the dark sector, and DM is also under its representation. To evade constraints from $Z'$ searches and other experiments at the LHC, the most straightforward strategy is to demand that all SM fermions carry no $\UoneX$ charges, which means $\UoneX$ is a hidden gauge symmetry~\cite{Langacker:2008yv, Pospelov:2007mp, Feldman:2007wj, Mambrini:2010dq}. In this scenario, the kinetic mixing between $\UoneX$ and $\UoneY$ bridges the dark sector with the SM sector. However, several studies have shown that such mixing is difficult to simultaneously generate dark matter relic abundance and satisfy the direct detection constraint~\cite{Chiu:1966kg, Kolb:1990vq, Gondolo:1990dk}, necessitating the introduction of additional parameters for a viable DM model. 

In this paper, we propose a $\UoneX$ DM model with kinetic and direct mixings, in which two new complex scalars are under both the $\UoneX$ and $\UoneY$ representations. Particularly, these mixings offer a plausible explanation for the W boson mass anomaly observed by the CDF Collaboration. We perform a parameter fitting utilizing data from both the CDF and non-CDF sources simultaneously. Furthermore, we investigate other constraints from colliders, including searches for $Z'$ and the $125~\mathrm{GeV}$ Higgs boson measurements. Additionally, we study comprehensively the phenomenology of the dark sector, particularly considering the direct detection experiments, the relic abundance of DM, and the indirect detection of DM annihilation. In calculating the constraint from the direct detection experiment, we consider the isospin violation in the scattering of DM and nucleons~\cite{Kang:2010mh, Chun:2010ve, Frandsen:2011cg, Gao:2011ka, Belanger:2013tla, Chen:2014tka}.

This paper is organized as follows. In Sec.~\ref{model}, we provide a detailed description of our $\UoneX$ model. In Sec.~\ref{constraints}, we discuss the phenomenology of this model regarding the electroweak precision measurements, $Z'$ searches at the LHC, Higgs physics, direct detection experiments, DM relic abundance, and indirect detection experiments. Sec.~\ref{Parameter scan} presents the results of a random scan in the parameter space. At last, we offer conclusions in Sec.~\ref{conclusions}.

\section{Model}
\label{model}

The model extends the SM with a new gauge symmetry $\UoneX$ and two complex scalars $S$ and $\Phi$. The $\Phi$ develops a large vacuum expectation value $v_\Phi$ in the early universe, and $S$ serves as the DM candidate. $S$ and $\Phi$ are both in the representation (1, $n_\Phi$, 1) of the gauge symmetry $\SUtwoL \times \UoneY \times \UoneX$. Consequently, the supercharges of $S$ and $\Phi$ provide a direct mixing between $\UoneY$ and $\UoneX$, which, along with the kinetic mixing, connects the dark sector with the SM particles.

\subsection{Lagrangian}
The full Lagrangian of the model is given by
\begin{eqnarray}\label{eq:Lag}
	\mathcal{L} &=&  \mathcal{L}_{SM} + (D^{\mu}S)^{\dag}(D_{\mu}S)+(D^{\mu}\Phi)^{\dag}(D_{\mu}\Phi) -\frac{1}{4}X^{\mu\nu}X_{\mu\nu} -\frac{s_{\varepsilon}}{2}B^{\mu\nu}X_{\mu\nu}
	\nonumber\\
	&&
	-\mu_{S}^{2}|S|^{2} - \lambda_{S}|S|^{4}
	+\mu_{\Phi}^{2}|\Phi|^{2}  - \lambda_{\Phi}|\Phi|^{4} -\lambda_{HS}|H|^{2}|S|^{2}
	-\lambda_{H\Phi}|H|^{2}|\Phi|^{2}
	-\lambda_{S\Phi}|S|^{2}|\Phi|^{2}.
\end{eqnarray}
Here the quadratic coefficient of DM is required to be negative, or it would become unstable by acquiring a non-zero vacuum expectation value. The covariant derivatives of $H$, $\Phi$ and $S$ are
\begin{eqnarray}\label{eq:covariant derivative}
D_\mu H &=& (\partial_\mu - ig T^a W^a_\mu  - i\frac{1}{2} g' B_\mu)H, \nonumber\\
D_\mu \Phi &=& (\partial_\mu - i g' n_\Phi B_\mu - ig_x X_\mu)\Phi, \nonumber\\
D_\mu S &=& (\partial_\mu - i g' n_\Phi B_\mu - ig_x X_\mu)S,
\end{eqnarray}
where $X_{\mu}$ is the new gauge boson of $\UoneX$ and $g_x$ is the gauge coupling. The scalar fields $H$ and $\Phi$ can be expanded in the unitary gauge as
\begin{eqnarray}
	H=\frac{1}{\sqrt{2}}
	\begin{pmatrix}
		0 \\
		v_H+h
	\end{pmatrix},\quad
	\Phi=\frac{1}{\sqrt{2}}(v_{\Phi}+\phi).
\end{eqnarray}
Then we can obtain the stationary point conditions by minimizing the scalar potential:
\begin{eqnarray}
\mu_{H}^{2} &=& \lambda_{H} v_H^{2} + \frac{1}{2} \lambda_{H\Phi}v_{\Phi}^{2} ,\nonumber\\
\mu_{\Phi}^{2} &=& \lambda_{\Phi}v_{\Phi}^{2} + \frac{1}{2}\lambda_{H\Phi} v_H^{2}.
\end{eqnarray}

The mass-squared matrix of real scalars ($h, \phi$) is given by
\begin{eqnarray}
\mathcal{L}^{\mathrm{mass}}_O &=& -\frac{1}{2}
\begin{pmatrix}
h & \phi
\end{pmatrix}
\begin{pmatrix}
2\lambda_{H}v_H ^{2}  & \lambda_{H\Phi}v_H v_{\Phi} \\
\lambda_{H\Phi}v_H v_{\Phi} & 2\lambda_{\Phi}v_{\Phi}^{2}
\end{pmatrix}
\begin{pmatrix}
h \\
\phi
\end{pmatrix} \nonumber\\
&=& -\frac{1}{2}
\begin{pmatrix}
	h_1 & h_2
\end{pmatrix}
\begin{pmatrix}
c_\theta & -s_\theta \\
s_\theta & c_\theta
\end{pmatrix}
\begin{pmatrix}
2\lambda_{H}v_H ^{2}  & \lambda_{H\Phi}v_H v_{\Phi} \\
\lambda_{H\Phi}v_H v_{\Phi} & 2\lambda_{\Phi}v_{\Phi}^{2}
\end{pmatrix}
\begin{pmatrix}
c_\theta & s_\theta \\
-s_\theta & c_\theta
\end{pmatrix}
\begin{pmatrix}
	h_1 \\
	h_2
\end{pmatrix} \nonumber\\
&=& -\frac{1}{2}
\begin{pmatrix}
h_1 & h_2
\end{pmatrix}
\begin{pmatrix}
m_{h_1}^2  & 0 \\
0 & m_{h_2}^2
\end{pmatrix}
\begin{pmatrix}
h_1 \\
h_2
\end{pmatrix},
\end{eqnarray}
where the rotation angle $\theta$ satisfies
\begin{eqnarray}
\tan 2\theta = \frac{\lambda_{H\Phi}v_H v_{\Phi}}{\lambda_{\Phi}v_{\Phi}^{2} - \lambda_{H}v_H ^{2}}.
\end{eqnarray}
Here we define $h_1$ is the $125~\mathrm{GeV}$ SM-like Higgs, and we choose $m_{h_2}$ as free parameter instead of $\lambda_\Phi$. Then we can express parameters $\lambda_H$ and $\lambda_\Phi$ by using physical observables $m_{h_1}$ and $m_{h_2}$:
\begin{eqnarray}
\lambda_H &=& \frac{m_{h_1}^2+m_{h_2}^2-\sqrt{(m_{h_1}^2+m_{h_2}^2)^2-4(m_{h_1}^2 m_{h_2}^2 + \lambda_{H\Phi}^2 v_H^2 v_\Phi^2)}}{4 v_H^2},\nonumber\\
\lambda_\Phi &=& \frac{m_{h_1}^2+m_{h_2}^2 + \sqrt{(m_{h_1}^2+m_{h_2}^2)^2-4(m_{h_1}^2 m_{h_2}^2 + \lambda_{H\Phi}^2 v_H^2 v_\Phi^2)}}{4 v_\Phi^2}.
\label{mass:higgs}
\end{eqnarray}

The DM mass $m_S$ can be obtained through the quadratic term of $S$. Similarly, we regard it as a free parameter, thereby expressing parameter $\mu_S^2$ as
\begin{eqnarray}
\mu_S^2 = m_S^2- \frac{1}{2} \lambda_{HS} v_H^2 - \frac{1}{2} \lambda_{S\Phi} v_\Phi^2.
\label{DM:stability}
\end{eqnarray}

\subsection{Gauge sector}
From the covariant derivatives of $H$, $\Phi$ and $S$, we can obtain the mass-squared matrix of gauge bosons ($W^3_\mu, B_\mu, X_\mu$) as
\begin{eqnarray}
&&\frac{1}{2}\frac{v_{H}^2}{4}(-gW_{\mu}^{3}+g' B_{\mu})^2+\frac{v_{\Phi}^2}{2}(g' n_{\Phi}B_{\mu}+g_{x}X_{\mu})^2 \nonumber\\
&=& \frac{1}{2}
\begin{pmatrix}
W^{3,\mu} & B^\mu & X^\mu
\end{pmatrix}
\begin{pmatrix}
\frac{g^2 v_H^2}{4} & -\frac{g g' v_H^2}{4} & 0\\
-\frac{g g' v_H^2}{4} & \frac{g'^2 v_H^2}{4} + g'^2 n_\Phi^2 v_\Phi^2 & g_x g' n_\Phi v_\Phi^2\\
0 & g_{x}g^{\prime}n_{\Phi}v_{\Phi}^2 & g_{x}^2v_{\Phi}^2
\end{pmatrix} 
\begin{pmatrix}
W^3_\mu\\
B_\mu\\
X_\mu
\end{pmatrix} \nonumber\\
&\equiv& \frac{1}{2}
\begin{pmatrix}
W^{3,\mu} & B^\mu & X^\mu
\end{pmatrix}
\mathcal{M}_g^2
\begin{pmatrix}
W^3_\mu\\
B_\mu\\
X_\mu
\end{pmatrix}.
\end{eqnarray}
Considering the kinetic mixing term $(s_\epsilon/2) B^{\mu \nu} X_{\mu \nu}$, we can eliminate it by a linear transformation~\cite{Babu:1997st},
\begin{eqnarray}
\begin{pmatrix}
W^3_\mu\\
B_\mu\\
X_\mu
\end{pmatrix} =
\begin{pmatrix}
1 & 0 & 0\\
0 &1 & -t_\epsilon\\
0 & 0 & 1/c_\epsilon
\end{pmatrix} 
\begin{pmatrix}
W^3_\mu\\
B'_\mu\\
X'_\mu
\end{pmatrix} \equiv \mathcal{K}
\begin{pmatrix}
W^3_\mu\\
B'_\mu\\
X'_\mu
\end{pmatrix},
\end{eqnarray}
where $c_\epsilon = \sqrt{ 1- s_\epsilon^2}$ and $t_\epsilon = s_\epsilon/c_\epsilon$. Then, the kinetic terms of the gauge bosons are canonically diagonalized in the ($W^3_\mu, B'_\mu, X'_\mu$) basis.
Then the mass-squared matrix $\mathcal{M}_g^2$ can be diagonalized by an orthogonal transformation $m_g^2 = O^T M_g^2 O = \mathrm{diag}\{0, m_Z^2, m_{Z'}^2\}$, where
\begin{eqnarray}
O &=& 
\mathcal{K}
\begin{pmatrix}
	1 & 0 & 0\\
	0 & c_\alpha & s_\alpha \\
	0 & -s_\alpha & c_\alpha
\end{pmatrix}
\begin{pmatrix}
	s_w' & c_w' & 0\\
	c_w' & -s_w' & 0 \\
	0 & 0 & 1
\end{pmatrix}
\begin{pmatrix}
	1 & 0 & 0\\
	0 & c_\xi & s_\xi \\
	0 & -s_\xi & c_\xi
\end{pmatrix} \\
&=&
\begin{pmatrix}
	s_w' & c_w' c_\xi & c_w' s_\xi\\
	c_w'(c_\alpha+s_\alpha t_\epsilon) & -s_\xi(s_\alpha-c_\alpha t_\epsilon)-c_\xi s_w'(c_\alpha+s_\alpha t_\epsilon) & c_\xi(s_\alpha-c_\alpha t_\epsilon)-s_\xi s_w'(c_\alpha+s_\alpha t_\epsilon) \\
	-\frac{c_w' s_\alpha}{c_\epsilon} & \frac{c_\xi s_\alpha s_w' - c_\alpha s_\xi}{c_\epsilon} & \frac{s_\xi s_\alpha s_w' + c_\alpha c_\xi}{c_\epsilon} \nonumber
\end{pmatrix},
\end{eqnarray}
here we have defined,
\begin{eqnarray}
n_x' &=& \frac{1}{c_\epsilon}-\frac{g'}{g_x}n_\Phi t_\epsilon,\\
t_\alpha &=& \frac{g' n_\Phi}{g_x n_x'},\\
g'' &=& g'(c_\alpha + t_\epsilon s_\alpha),\\
s_w' &=& \frac{g''}{\sqrt{g^2 + g''^2}},\\
c_w' &=& \frac{g}{\sqrt{g^2 + g''^2}},\\
t_\epsilon' &=& -\frac{s_\alpha - t_\epsilon c_\alpha}{c_\alpha + t_\epsilon s_\alpha},\\
n_x'' &=& \frac{g'}{g_x}n_\Phi s_\alpha + n_x'c_\alpha, \\
t_{2\xi} &=& \frac{2g'' t_\epsilon' v_H^2 \sqrt{g^2 + g''^2}}{v_H^2 (g^2 + g''^2) - g''^2 t_\epsilon'^2 v_H^2 - 4g_x^2 n_x''^2 v_\Phi^2}.
\end{eqnarray}
Finally, we obtain the physical gauge fields $(A_\mu, Z_\mu, Z'_\mu)$ by canonical normalizing the kinetic mixing term and diagonalizing the mass-squared matrix at the same time. The masses for $Z$ and $Z'$ are given by
\begin{eqnarray}
m_Z^2 &=& \frac{g^2 + g''^2}{4}v_H^2(1 + s_w' t_\epsilon' t_\xi),\\
m_{Z'}^2 &=& \frac{g_x^2 n_x''^2 v_\Phi^2}{1 + s_w' t_\epsilon' t_\xi}. \label{mass:mZp}
\end{eqnarray}
For the convenience of discussing the phenomenologies later, we show the neutral current interactions in terms of $(A_\mu, Z_\mu, Z'_\mu)$ here,
\begin{eqnarray}
	\mathcal{L}_{NC} = e A_\mu J_{EM}^\mu + Z_\mu \bar{f} \gamma^\mu (g^f_V - g^f_A \gamma^5)f + Z_\mu' \bar{f} \gamma^\mu (g'^f_V - g'^f_A \gamma^5)f,
\end{eqnarray}
where
\begin{eqnarray}\label{current}
	g^f_V &=& \frac{T^3_f}{2} \Big[g O_{12} - g' O_{22} \Big] + g'Q_f O_{22}, \nonumber\\
	g^f_A &=& \frac{T^3_f}{2}\Big[g O_{12} - g' O_{22} \Big], \nonumber\\
	g'^f_V &=& \frac{T^3_f}{2} \Big[g O_{13} - g' O_{23} \Big] + g'Q_f O_{23}, \nonumber\\
	g'^f_A &=& \frac{T^3_f}{2} \Big[g O_{13} - g' O_{23} \Big] .
\end{eqnarray}
Here the electromagnetic current $J_{EM}^\mu$ remains unchanged as the SM form, while the eletric charge reads $e=gO_{11} = g'O_{21}$.

We choose the set of free parameters in our model as
\begin{eqnarray}
\{ m_{h_2}, m_S, \lambda_S, \lambda_{HS}, \lambda_{H\Phi}, \lambda_{S\Phi}, g_x, n_\Phi, s_\epsilon, m_{Z'} \},\label{free parameters}
\end{eqnarray}
where the last four parameters are related to the gauge sector. And we have chosen $m_{Z'}$ as the free parameter instead of $v_\Phi$. 

\section{Constraints}
\label{constraints}

\subsection{Electroweak precision measurements}
In this subsection, we will explore whether our model can account for the recently observed $W$ boson mass.  Within our model framework, the couplings between $Z'$ and SM gauge bosons will lead to the deviation of $m_W$. Alternatively, this deviation can be interpreted as an additional contribution to the electroweak oblique parameters $S$, $T$, and $U$~\cite{Peskin:1990zt, Kumar:2013yoa}. Given the considerable difference in measurements between CDF and ATLAS, we fit our model parameters by utilizing the global fit results of $S$, $T$, and $U$ derived from both the CDF~\cite{deBlas:2022hdk} and non-CDF sources~\cite{Haller:2018nnx}. Then we compare their fitting results. Table~\ref{tab:STU} shows the global fit results of $S$, $T$, and $U$ under these two scenarios.
\begin{table}[htb]
\centering
\resizebox{0.75\textwidth}{!}{
\begin{tabular}{|c|c|rrr|c|rrr|}
\hline
&CDF results & \multicolumn{3}{c|}{Correlation} &Non-CDF results & \multicolumn{3}{c|}{Correlation} \\\hline
$S$ & $ \phantom{+}0.005 \pm 0.096 $ & \phantom{+}$1.00$ & & & $ \phantom{+}0.040 \pm 0.110 $ & \phantom{+}$1.00$ & & \\ \hline
$T$ & $ \phantom{+}0.040 \pm 0.120 $ & $0.91$ & $1.00$ & & $ \phantom{+}0.090 \pm 0.140 $ & $0.92$ & $1.00$ & \\ \hline
$U$ & $ \phantom{+}0.134 \pm 0.087 $ & $-0.65$ & $-0.88$ &$1.00$ & $ -0.020 \pm 0.110 $ & $-0.68$ & $-0.87$ & $1.00$ \\ \hline
\end{tabular}}
\caption{Results of the global fit of the oblique parameters by fitting the CDF and non-CDF sources.}
\label{tab:STU}
\end{table}

Using the effective Lagrangian techniques, the electroweak charged and neutral current interaction can be written as related to the electroweak oblique parameters~\cite{Burgess:1993vc}:
\begin{eqnarray}
\mathcal{L}_{CC, W} &=& -\frac{e}{\sqrt{2} \hat{s}_{w}}(1-\frac{\alpha S}{4(\hat{c}_{w}^2-\hat{s}_{w}^2)}+\frac{\hat{c}_{w}^2\alpha T}{2(\hat{c}_{w}^2-\hat{s}_{w}^2)}+\frac{\alpha U}{8\hat{s}_{w}^2})\sum\limits_{ij}V_{ij}\bar{f}_{i}\gamma^{\mu}\gamma_{L}f_{j}W_{\mu}^{\dagger}+\mathrm{c.c.}, \nonumber\\  
\mathcal{L}_{NC, Z} &=& \frac{e}{\hat{s}_{w}\hat{c}_{w}}(1+\frac{\alpha T}{2})\sum\limits_{f}\bar{f}\gamma^{\mu}[T^{3}_{f}\frac{1-\gamma^{5}}{2}-Q_{f}(\hat{s}_{w}^2+\frac{\alpha S}{4(\hat{c}_{w}^2-\hat{s}_{w}^2)}-\frac{\hat{c}_{w}^2\hat{s}_{w}^2\alpha T}{\hat{c}_{w}^2-\hat{s}_{w}^2})]f Z_{\mu}.
\end{eqnarray}
Here $Q_f$ denotes the electric charge, defined by $Q_f = Y_f + T^3_f$. $Y_f$ is the hypercharge and $T^3_f$ is the third component of weak isospin. It should be noted that $\hat{s}_w$ and $\hat{c}_w$ with a hat overhead are physical observables, defined by 
\begin{eqnarray}\label{eq:covariant derivative}
\frac{G_F}{\sqrt{2} } = \frac{4 \pi \alpha }{8 \hat{s}_w^2 \hat{c}_w^2 m_Z^2}.
\end{eqnarray}
The charged current interaction of our model remains the same as SM. In contrast, the neutral current interaction is modified as Eq.~(\ref{current}).
Note that the parameters $g$ and $g'$ with no hat overhead are unphysical and dependent upon the input value of other free parameters. Then we can obtain the parameter relationships between $S$, $T$, $U$ and our model:
\begin{eqnarray}
\alpha T &=& 2\hat{c}_w \hat{s}_w \frac{-g' O_{22} + g O_{12} }{e} - 2, \nonumber\\
\alpha S &=& \frac{-4g' O_{22} (\hat{c}_w^2 - \hat{s}_w^2)}{-g' O_{22} + g O_{12}}  - 4\hat{s}_w^2(\hat{c}_w^2 - \hat{s}_w^2) + 4\hat{c}_w^2 \hat{s}_w^2 \alpha T, \nonumber\\
\alpha U &=& 8\hat{s}_w^2(\frac{\hat{s}_w}{s_w} - 1 + \frac{\alpha S}{4(\hat{c}_w^2 - \hat{s}_w^2)} - \frac{\hat{c}_w^2 \alpha T}{2(\hat{c}_w^2 - \hat{s}_w^2)} ).
\end{eqnarray}

\begin{figure}[!h]
\centering
\subfigure[\label{fig1-1}]
{\includegraphics[width=0.48\textwidth]{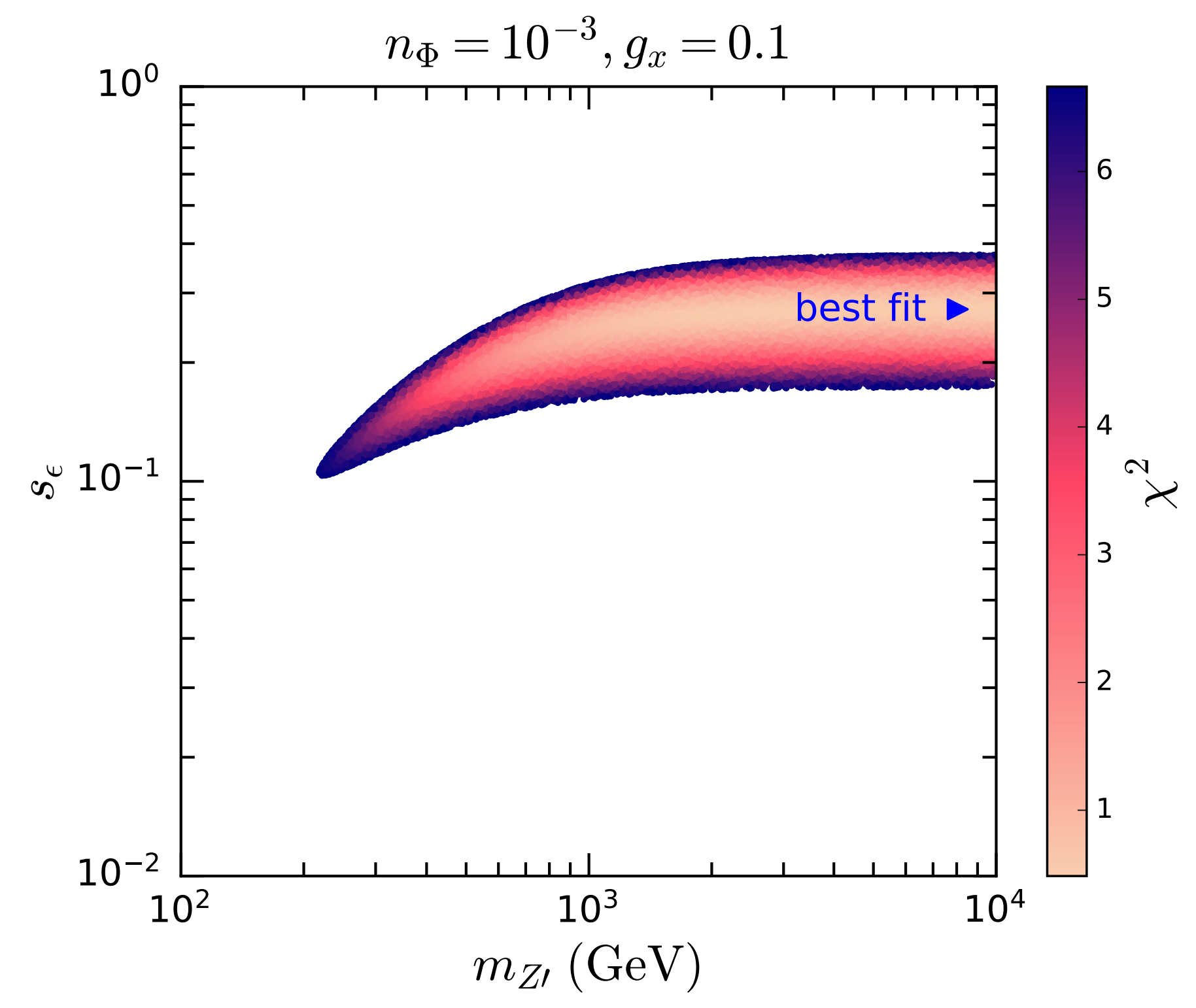}}
\hspace{.01\textwidth}
\subfigure[\label{fig1-2}]
{\includegraphics[width=0.48\textwidth]{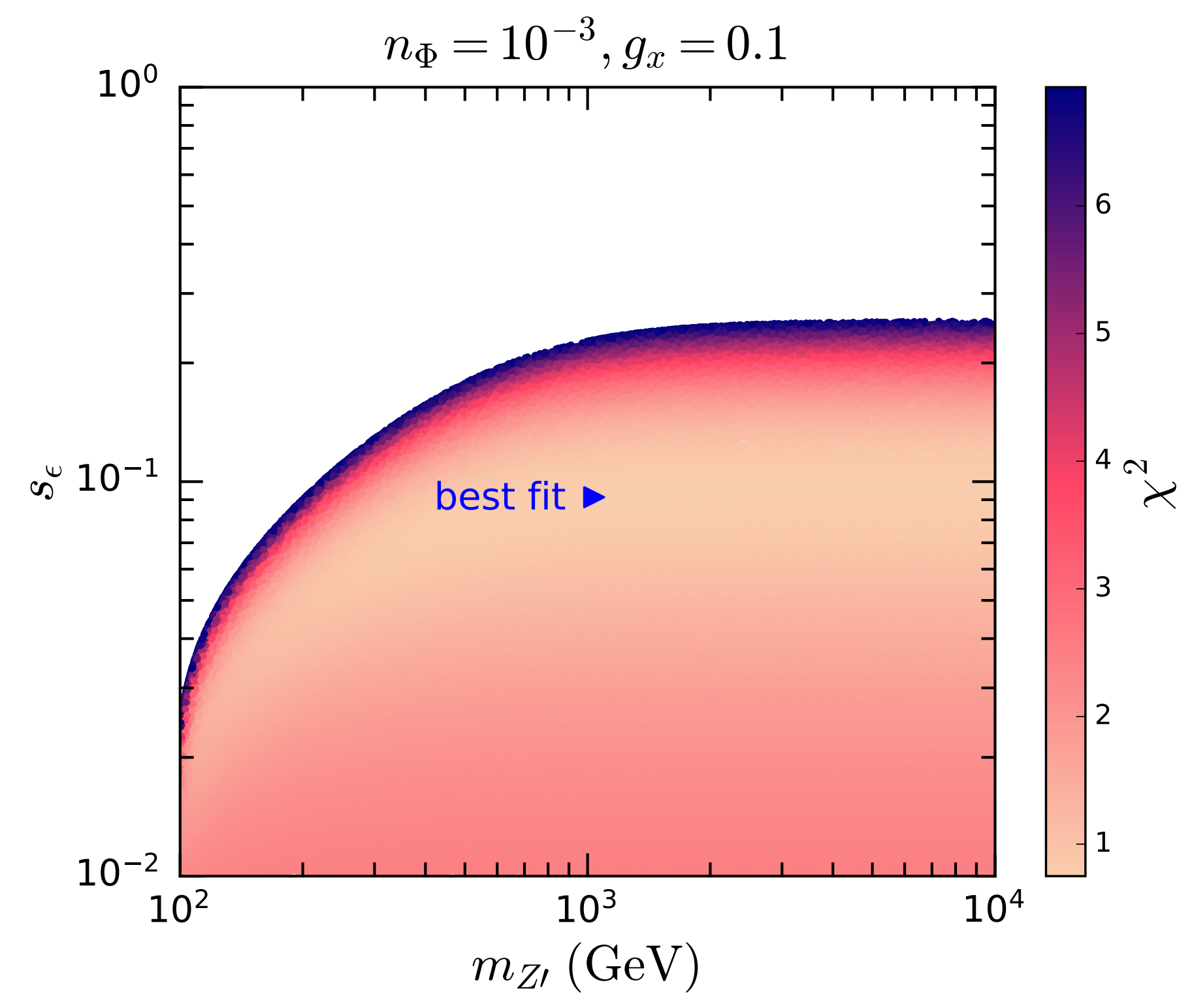}}
\hspace{.01\textwidth}
\subfigure[\label{fig1-3}]
{\includegraphics[width=0.48\textwidth]{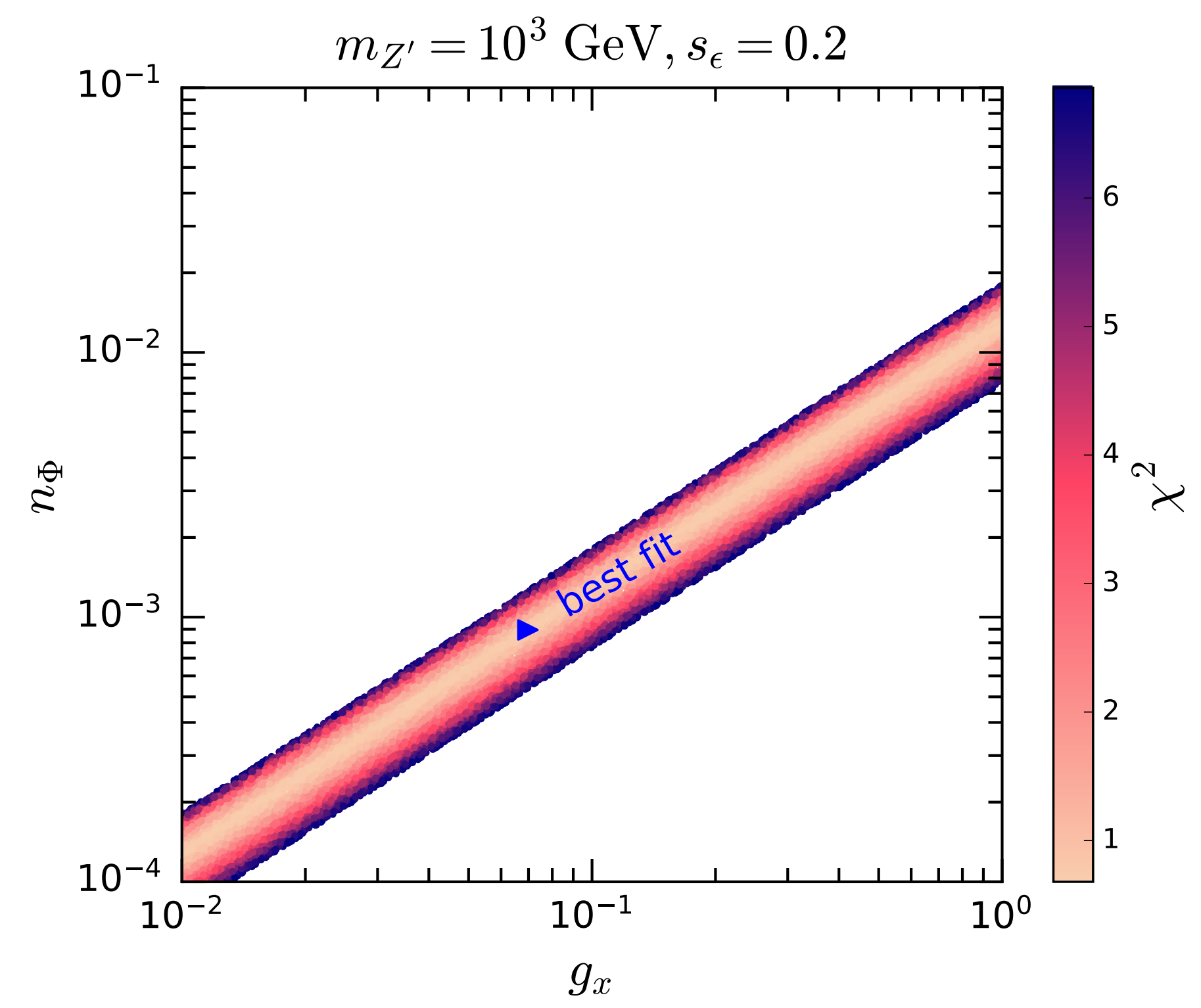}}
\subfigure[\label{fig1-4}]
{\includegraphics[width=0.48\textwidth]{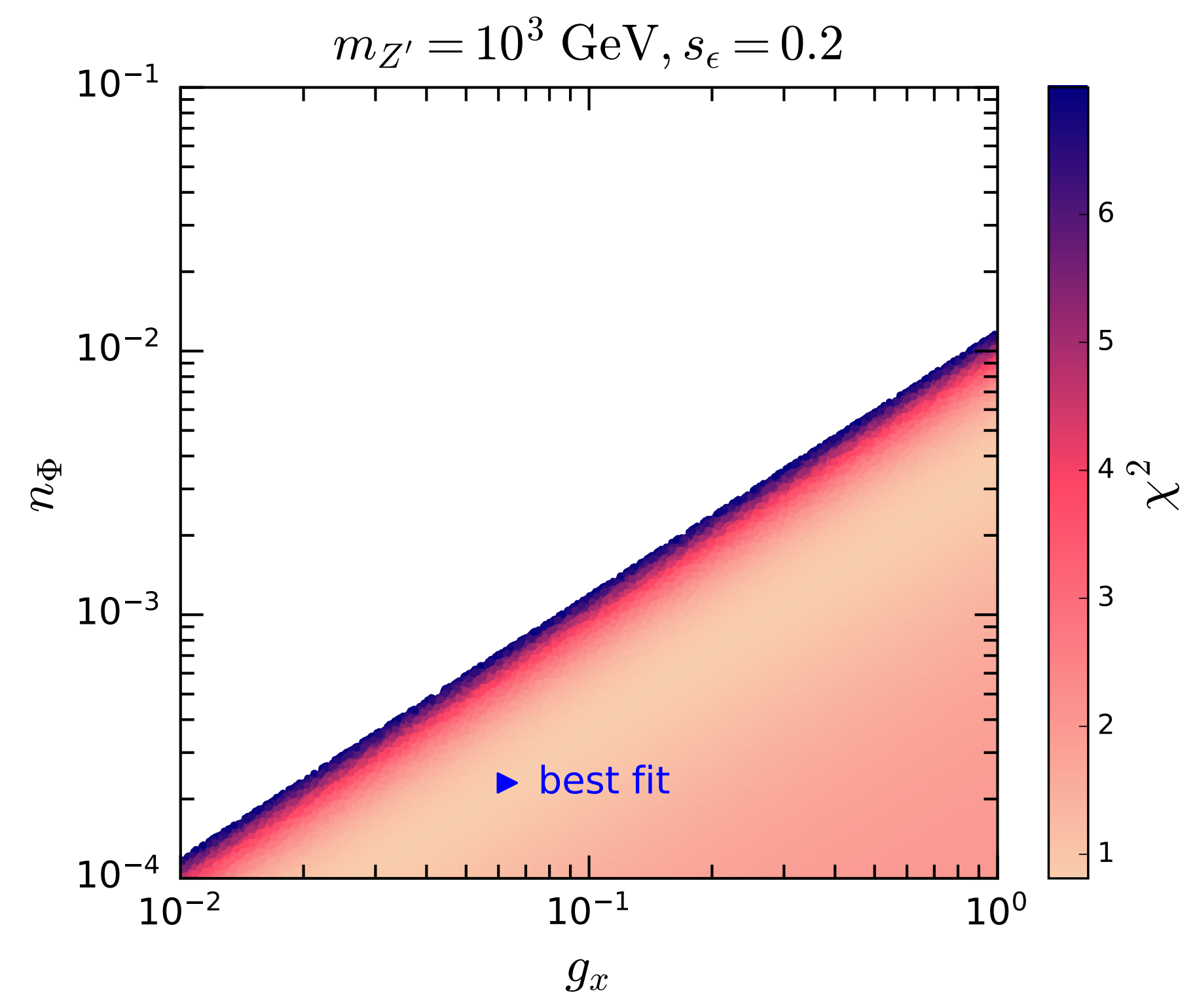}}
\caption{Fitting results of the model parameters in the $s_{\epsilon}$-$m_{Z'}$ plane (Top) and $n_\Phi$-$g_x$ plane (Bottom) by fitting CDF measurement (Left) and non-CDF measurement (Right). The blue triangle gives the best fit of electroweak oblique parameters, and the colored region corresponds to $\Delta \chi^2 = 6.18$ with respect to the best fit.}
\label{fig1}
\end{figure}

Based on the calculations detailed in Sec.~\ref{model}, the free parameters associated with $S$, $T$, $U$ in our model include $g_x$, $n_\Phi$, $s_\epsilon$ and $m_{Z'}$. In Fig.~\ref{fig1}, we plot the fitting results of these free parameters. The blue triangle represents the best fit point in each figure and we utilize gradient colors to fill the region where $\Delta \chi^2 = 6.18$ relative to the best fit value. Additionally, the $\chi^2$ value of each point is visualized through the colorbar on the right side of each plot. Figs.~\ref{fig1-1} and \ref{fig1-3} on the left depict the fitting results based on CDF measurement, while Figs.~\ref{fig1-2} and \ref{fig1-4} on the right correspond to fitting non-CDF measurement. Moreover, Figs.~\ref{fig1-1} and \ref{fig1-2} at the top display the results in the $s_{\epsilon}$-$m_{Z'}$ plane with the other parameters fixed to be $n_\Phi = 10^{-3}$ and $g_x = 0.1$, while Figs.~\ref{fig1-3} and \ref{fig1-4} at the bottom present the fitting results in the $n_\Phi$-$g_x$ plane with the fixed parameters chosen to be $m_{Z'} = 10^3~\mathrm{GeV}$ and $s_\epsilon = 0.2$. The parameter values corresponding to each best fit point are provided in in Table~\ref{tab:bestfit}.

\begin{table}[htb]
\centering
\resizebox{0.48\textwidth}{!}{
\begin{tabular}{|c|c|c|c|c|c|}
\hline
&$g_x$ & $n_\Phi$ & $s_\epsilon$ & $m_{Z'}(\mathrm{GeV})$ & $\chi^2$ \\\hline
Fig.~\ref{fig1-1} & $0.1$ & $10^{-3}$  & $0.27$ & $8047.6$ & 0.48 \\ \hline
Fig.~\ref{fig1-2} & $0.1$ & $10^{-3}$  & $0.09$ & $1037.8$ & 0.75\\ \hline
Fig.~\ref{fig1-3} & $0.070$ & $8.9 \times 10^{-4}$  & $0.2$ & $1000$ & 0.68 \\ \hline
Fig.~\ref{fig1-4} & $0.063$ & $2.3 \times 10^{-4}$  & $0.2$ & $1000$ & 0.81\\ \hline
\end{tabular}}
\caption{Parameter Values of best fit points in Fig.~\ref{fig1}.}
\label{tab:bestfit}
\end{table}

Due to the significant discrepancies in measurements between CDF and other experiments,  the fitting results for model parameters diverge substantially in these two scenarios. The CDF measurement exhibits a $7\sigma$ deviation from the SM prediction, suggesting that the interaction between $Z'$ and SM gauge bosons must fall within a specific range. Specifically,  as depicted in Fig.~\ref{fig1-1}, $m_{Z'}$ is greater than around $200~\mathrm{GeV}$ and $s_\epsilon$ lies within $0.1 \sim 0.4$. Meanwhile, Fig.~\ref{fig1-3} shows a linear correlation between $g_x$ and $n_\Phi$, arising from the fact that $\Phi$ possesses both $\UoneY$ and $\UoneX$ charges. As $g_x$ increases, the couplings between $Z'$ and SM particles weaken, necessitating an increase in $n_\Phi$. By comparison, in the case of fitting non-CDF measurement, the interaction between $Z'$ and SM particles must be sufficiently small. Notably, both $s_\epsilon$ and $n_\Phi$ exhibit upper bounds, which vary with $m_{Z'}$ and $g_x$ respectively.

\subsection{$Z'$ searches at the LHC}
The couplings between $Z'$ and SM fermions can induce the production of $Z'$ in the s-channel at hadron colliders, leading to its detection through the decay products $Z' \rightarrow l^+ l^-$ ($l \equiv e, \mu, \tau$). The ATLAS and CMS collaborations routinely conduct searches for the $Z'$ boson at the LHC~\cite{ATLAS:2019erb, CMS:2021ctt}. Since no new particles have been discovered so far, experiments still impose stringent upper limits upon the cross section for $Z'$ production.
	
Using \texttt{FeynRules}~\cite{Alloul:2013bka} and \texttt{MadGraph5\_aMG@NLO}~\cite{Alwall:2014hca}, we calculate the cross section $\sigma(p p \to Z' X \to e^+ e^- X)$ for producing $Z'$ at the LHC, which subsequently decays to the $e^+ e^-$ final state. We then compare this result with the $95\%$ confidence level (CL) upper limit dilepton data set observed by ATLAS at the $13~\mathrm{TeV}$ LHC~\cite{ATLAS:2019erb}.

\begin{figure}[!h]
\centering
\subfigure[\label{fig2-1}]
{\includegraphics[width=0.48\textwidth]{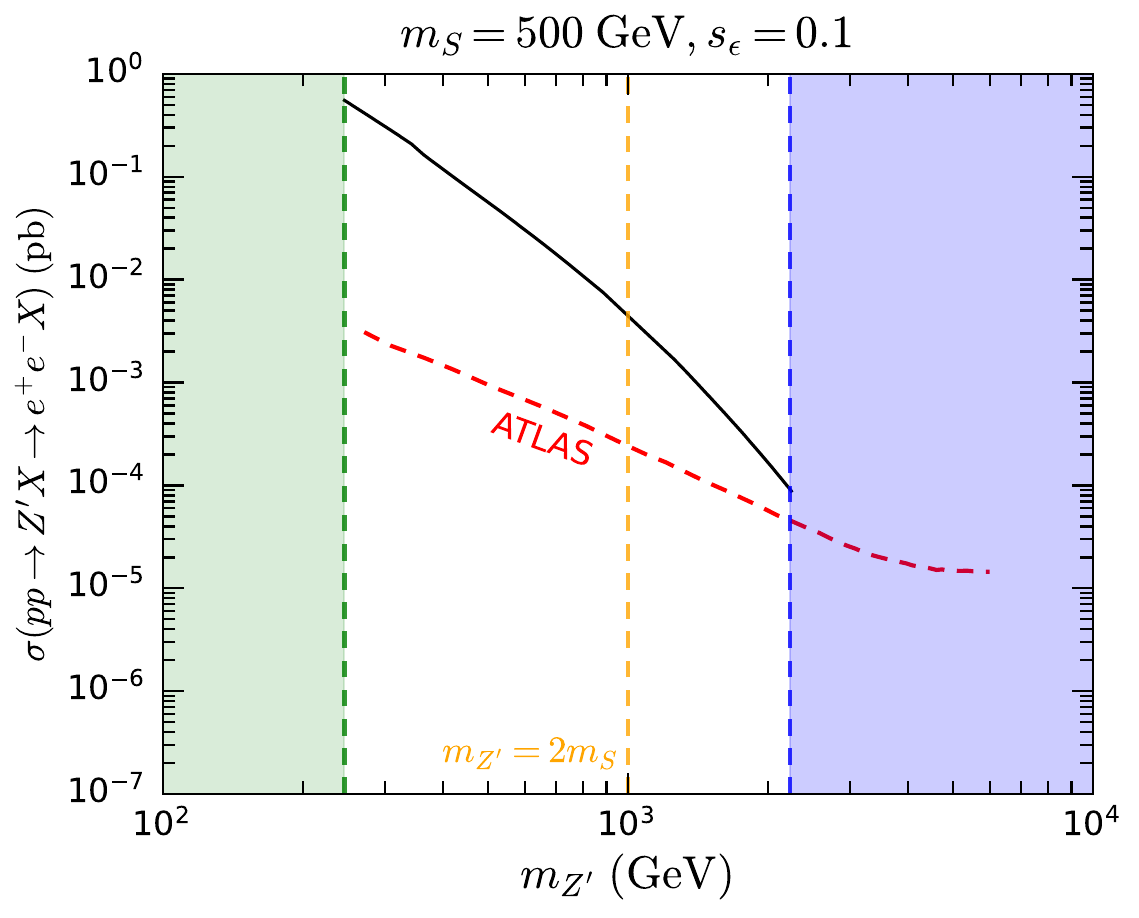}}
\hspace{.01\textwidth}
\subfigure[\label{fig2-2}]
{\includegraphics[width=0.48\textwidth]{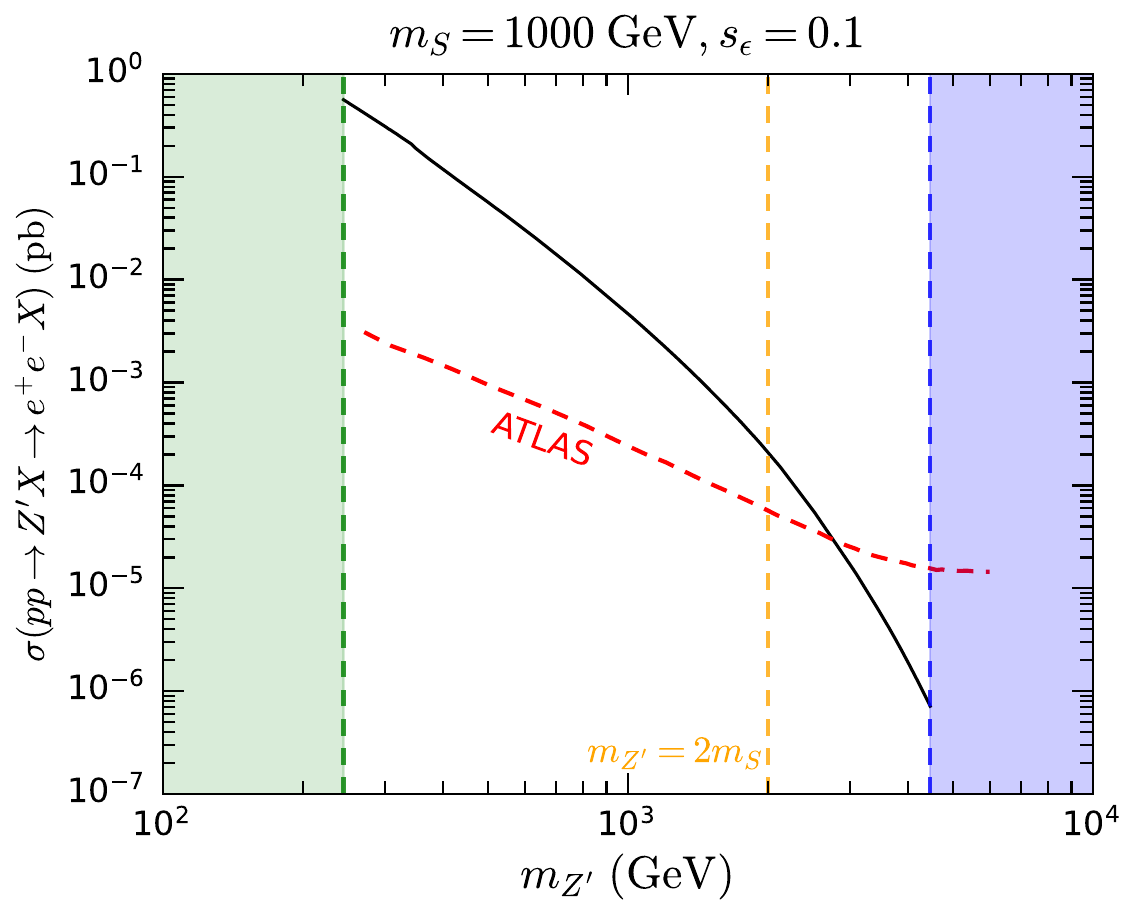}}
\hspace{.01\textwidth}
\subfigure[\label{fig2-3}]
{\includegraphics[width=0.48\textwidth]{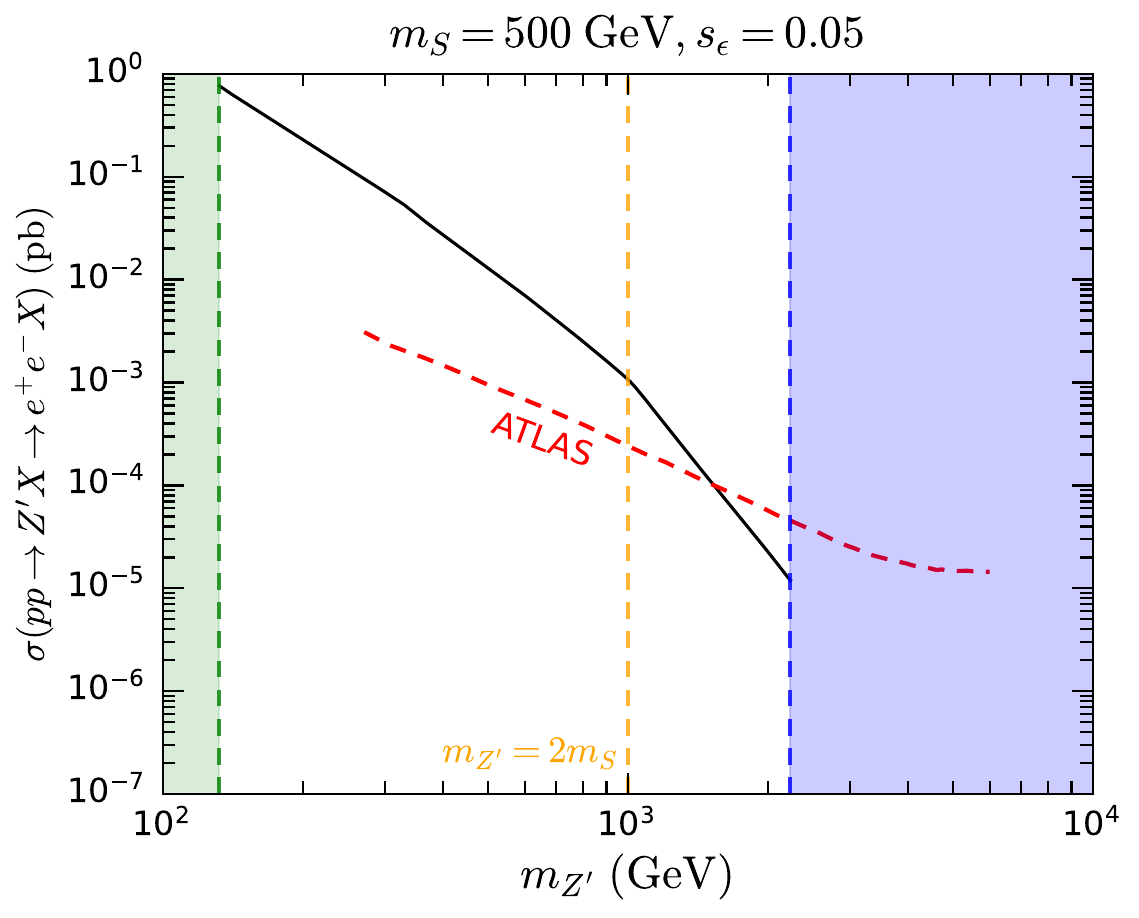}}
\subfigure[\label{fig2-4}]
{\includegraphics[width=0.48\textwidth]{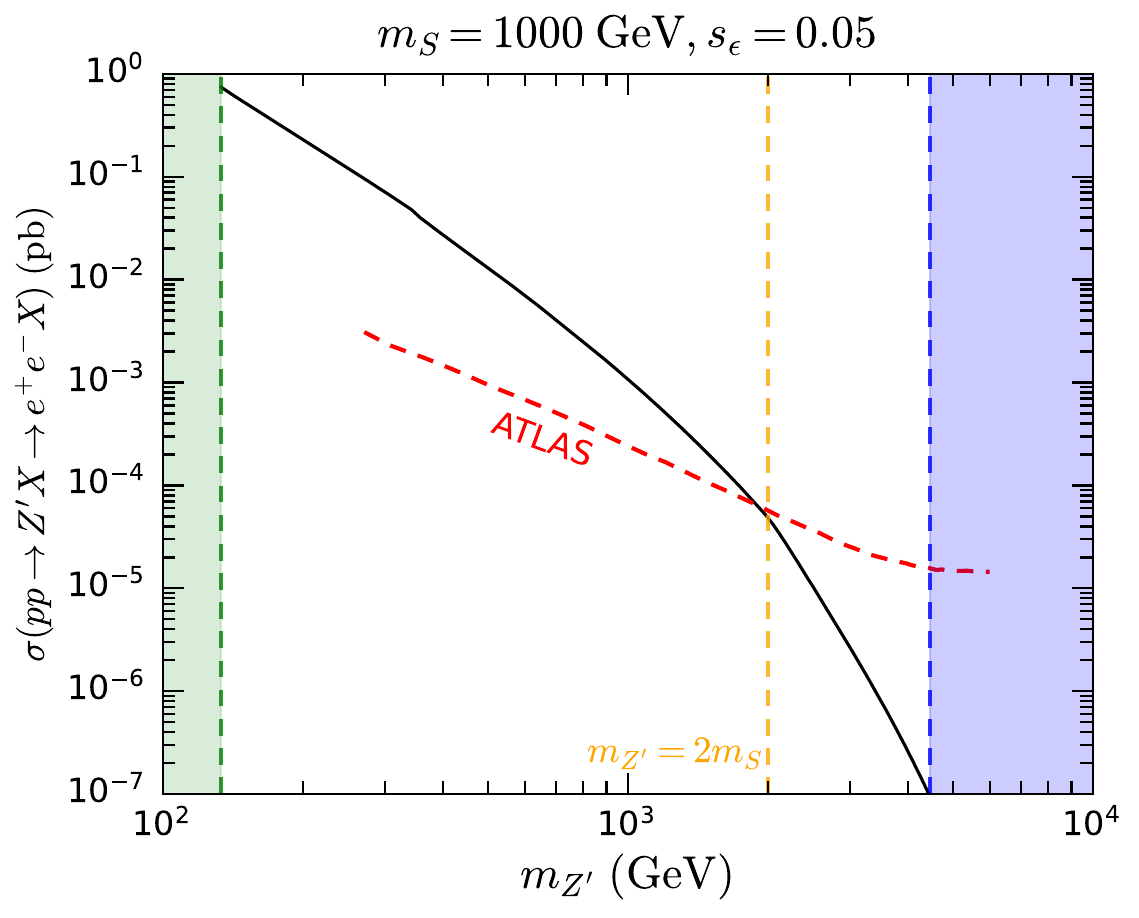}}
\caption{Cross section $\sigma(p p \to Z' X \to e^+ e^- X)$ as a function of $m_Z'$ for $s_\epsilon = 0.1$ (Top) and $0.05$ (Bottom) with $m_S = 500~\mathrm{GeV}$ (Left) and $1000~\mathrm{GeV}$ (Right). The red dashed line denotes the $95\%$ CL upper limit set by ATLAS, and the orange vertical line corresponds to $m_{Z'} = 2m_S$. The green area lies outside the $\Delta \chi^2 = 6.18$ region in Fig.~\ref{fig1-2} while the blue area is excluded due to the stability of DM.}
\label{fig2}
\end{figure}

In Fig.~\ref{fig2}, we select different $m_S$ and $s_\epsilon$ as benchmark points with other relevant parameters to be $\lambda_S = \lambda_{HS} = \lambda_{H\Phi} = \lambda_{S\Phi} = 10^{-3}$, $m_{h_2} = 1~\mathrm{TeV}$, $n_\Phi = 10^{-3}$ and $g_x = 0.1$. Figs.~\ref{fig2-1} and \ref{fig2-2} at the top  correspond to the horizontal line $s_\epsilon = 0.1$ in Fig.~\ref{fig1-2} for $m_S = 500~\mathrm{GeV}$ and $1000~\mathrm{GeV}$ respectively. In contrast, Figs.~\ref{fig2-3} and \ref{fig2-4} at the bottom correspond to the case for $s_\epsilon = 0.05$. The green area in Fig.~\ref{fig2} lies outside the $\Delta \chi^2 = 6.18$ region in Fig.~\ref{fig1-2}. It's worth noting that using the fitting result in Fig.~\ref{fig1-1} will alter the green area, as it depends on the electroweak oblique parameters in the fitting process. Nevertheless, the analysis remains similar. The blue area is excluded by ensuring the stability of DM, which can be determined through Eq.~(\ref{DM:stability}). The black line represents the 
$Z'$ production cross section while the red dashed line indicates the $95\%$ CL upper limit set by ATLAS.

In Fig.~\ref{fig2-1}, the entire $m_{Z'}$ region is excluded. However, in Fig.~\ref{fig2-3}, as $s_\epsilon$ decreases to $0.05$, the $m_{Z'}$ region survives from $1527~\mathrm{GeV}$ to $2245~\mathrm{GeV}$ due to the weakening interaction between $Z'$ and SM particles. Similarly, in Fig.~\ref{fig2-2} as $m_S$ increases to $1000~\mathrm{GeV}$, the surviving $m_{Z'}$ region extends from $2754~\mathrm{GeV}$ to $4464~\mathrm{GeV}$ owing to the rise of the lower limit imposed by DM stability constraints. Additionally, when $m_{Z'} \geq 2m_S$, the cross section decreases significantly faster. This is attributed to the kinematically allowed decay channel $Z' \to S^\dagger S$, leading to a decrease in the branching fraction for $Z' \to e^+ e^-$. This effect becomes more pronounced with decreasing $s_\epsilon$, as evident in Figs.~\ref{fig2-3} and \ref{fig2-4}. Table~\ref{tab:zpsearch} presents the surviving $m_{Z'}$ regions depicted in each subfigure of Figure \ref{fig2}, which will be further investigated by future $Z'$ search experiments.

\begin{table}[htb]
\centering
\resizebox{0.38\textwidth}{!}{
\begin{tabular}{|c|c|c|c|}
\hline
&$m_S(\mathrm{GeV})$ & $s_\epsilon$ & $m_{Z'}(\mathrm{GeV})$  \\\hline
Fig.~\ref{fig2-1} & $500$ & $0.1$  &    \\ \hline
Fig.~\ref{fig2-2} & $1000$ & $0.1$  & $2754 \sim 4464$ \\ \hline
Fig.~\ref{fig2-3} & $500$ & $0.05$&  $1527 \sim 2245$  \\ \hline
Fig.~\ref{fig2-4} & $1000$ &$0.05$&  $1908 \sim 4507$  \\ \hline
\end{tabular}}
\caption{Surviving region of $m_{Z'}$ in each subfigure of Fig.~\ref{fig2}.}
\label{tab:zpsearch}
\end{table}

\subsection{Higgs physics}
In this subsection, we discuss the implications upon the Higgs sector arising from the presence of new particles in our model. As a consequence of the mixing between the Higgs bosons $h_1$ and $h_2$, as well as between the gauge bosons $Z$ and $Z'$, the couplings between Higgs and SM particles deviate from their expected values. These deviations can be described by the following expression~\cite{LHCHiggsCrossSectionWorkingGroup:2013rie}
\begin{eqnarray}
\mathcal{L}_{h_1} \supset \kappa_W \frac{2m_W^2}{v_H}h_1 W^+_\mu W^{-,\mu} + \kappa_Z \frac{m_Z^2}{v_H}h_1 Z_\mu Z^{\mu}-\sum_f \kappa_f \frac{m_f}{v_H}h_1 \bar{f}f,
\end{eqnarray}
where
\begin{eqnarray}
\kappa_W &=& \kappa_f = c_\theta, \\
\kappa_Z &=& c_\theta (c_W O_{22} - s_W O_{12})^2 - (n_\Phi^2 g'^2 O_{22}^2 + 2n_\Phi g' g_x O_{22} O_{32} + g_x^2 O_{32}^2)\frac{s_\theta v_\Phi v_H}{m_Z^2}.
\end{eqnarray}
These couplings elucidate the modifications to the Higgs interactions induced by the incorporation of new particles. In SM, all these couplings equal to $1$. Moreover, if kinematically allowed, there may exist additional Higgs decay channels. If $m_{h_1} > 2 m_{h_2}$, the undetected decay channel $h_1 \to 2 h_2$ becomes accessible, with its decay width given by
\begin{eqnarray}
\Gamma(h_{1} \to 2h_2)=\frac{g_{h_{1} h_2 h_2}^2}{32\pi m_{h_1}}\sqrt{1-\frac{4m_{2}^2}{m_{h_1}^2}},
\end{eqnarray}
where
\begin{eqnarray}
g_{h_{1} h_2 h_2} = 3\lambda_{H}v_H c_\theta s_\theta^2 - 3\lambda_{\Phi}v_\Phi s_\theta c_\theta^2 + \frac{\lambda_{H\Phi}}{2} \Big[v_H(c_\theta^3 - 2c_\theta s_\theta^2) - v_\Phi(s_\theta^3 - 2c_\theta^2 s_\theta)) \Big].
\end{eqnarray}
Similarly, if $m_{h_1} > 2 m_S$, the invisible decay channel $h_1 \to S^\dagger S$ opens, with its decay width given by
\begin{eqnarray}
\Gamma(h_{1} \to S^\dagger S)=\frac{g_{h_{1} S^\dagger S}^2}{16\pi m_{h_1}}\sqrt{1-\frac{4m_{S}^2}{m_{h_1}^2}},
\end{eqnarray}
where
\begin{eqnarray}
g_{h_1S^\dagger S} = \lambda_{HS}v_H c_\theta - \lambda_{S\Phi}v_\Phi s_\theta.
\end{eqnarray}

We utilize a numerical tool \texttt{Lilith~2}~\cite{Bernon:2015hsa, Kraml:2019sis} to constrain our model parameter space in Sec.~\ref{Parameter scan}. This tool, implemented as a public Python library, facilitates the examination of constraints imposed on new physics models by Higgs signal strength measurements, including data from ATLAS and CMS Run 2 Higgs results for $36~\mathrm{fb}^{-1}$. To ensure compatibility with experimental observations at the $95\%$ CL, we exclude parameter points with p-values below $0.05$. This rigorous methodology enables us to effectively navigate the vast parameter space and identify regions consistent with experimental data.

\subsection{DM-nucleon scattering}
\begin{figure}[!h]
\centering
\subfigure[Vector interaction mediated by $Z$ and $Z'$.\label{ssffV}]
{\includegraphics[width=0.25\textwidth]{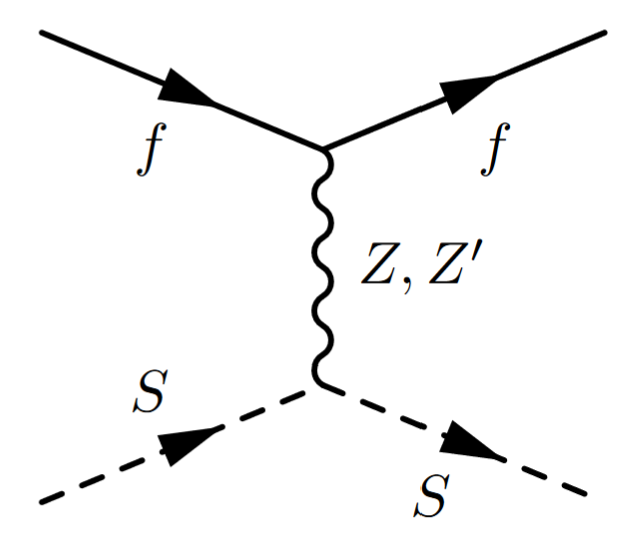}}
\hspace{.2\textwidth}
\subfigure[Scalar interaction mediated by $h_1$ and $h_2$.\label{ssffS}]
{\includegraphics[width=0.25\textwidth]{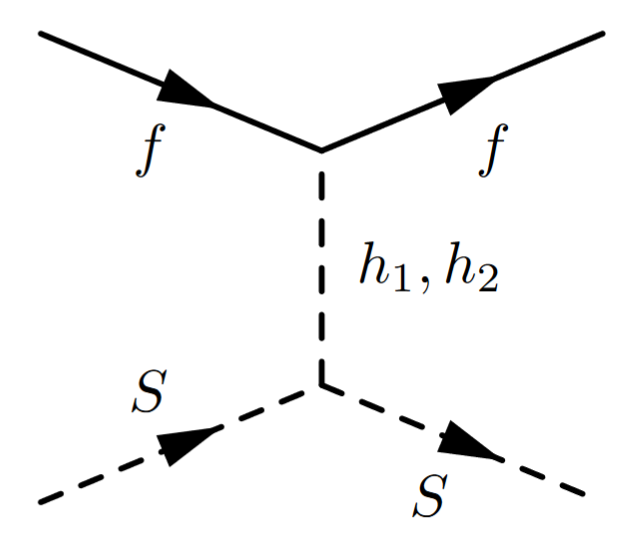}}
\caption{ DM-fermions scattering mediated by vector bosons (a) and scalars (b). }
\label{ssff}
\end{figure}

In this model, two types of interactions facilitate spin-independent (SI) DM-nucleon scattering. One is vector interactions, mediated by gauge bosons $Z$ and $Z'$. The other is scalar interactions, mediated by the scalars $h_1$ and $h_2$. Fig.~\ref{ssff} illustrates the Feynman diagrams corresponding to these two interaction types.

To derive the cross section, we express the scattering interaction between DM $S$ and SM fermions $f$ using the effective field theory presented in Ref.~\cite{Yu:2011by}:
\begin{eqnarray}
\mathcal{L}_S &=& \sum_f F_{S,f} S^\dagger S \bar{f} f, \nonumber\\
\mathcal{L}_{V} &=& \sum_f F_{V,f} (S^\dagger i \overleftrightarrow{\partial_\mu} S ) \bar{f} \gamma^\mu f.
\end{eqnarray}	
Here $F_{S,f}$ represents the effective coupling for the scalar interaction while $F_{V,f}$ corresponds to the vector interaction. Note that the axial vector interaction can be neglected due to the velocity suppression in the low-velocity limit~\cite{Yu:2011by}. To compute the cross section, we define four effective couplings $F_{S, N}$ and $F_{V,N}$ ($N = p$, $n$). $F_{S, N}$ are given by
\begin{eqnarray}
F_{S,N} &=& \sum_{q = u,d,s} F_{S,q} f_q^N \frac{m_N}{m_q} + \sum_{q = c,b,t} F_{S,q} f_Q^N \frac{m_N}{m_q} \\
&=& \frac{F(\theta)}{9} m_N [2 + 7(f_u^N + f_d^N + f_s^N)],
\end{eqnarray}
with
\begin{eqnarray}
F(\theta) = \frac{\lambda_{HS} (c_\theta^2 m_{h_2}^2 + s_\theta^2  m_{h_1}^2)}{ m_{h_1}^2  m_{h_2}^2} - \frac{\lambda_{S\Phi}v_\Phi s_\theta c_\theta (  m_{h_2}^2 -  m_{h_1}^2)}{v_H  m_{h_1}^2  m_{h_2}^2}.
\end{eqnarray}
Here, $\theta$ is the mixing angle between $h_1$ and $h_2$. $f_q^N$ are the nucleon form factors for each light quark, and $f_Q^N = \frac{2}{27}(1 - f_u^N - f_d^N - f_s^N)$ for heavy quarks~\cite{Ellis:2000ds}. $F_{V,N}$ are given by
\begin{eqnarray}
F_{V,p} = 2F_{V,u} + F_{V,d}, \quad F_{V,n} = F_{V,u} + 2F_{V,d},
\end{eqnarray}
with
\begin{eqnarray}
F_{V,f} &=& g_V^f  \frac{\mathcal{KO}_{32}}{g_x v_\Phi^2} + g_V'^f \frac{\mathcal{KO}_{33}}{g_x v_\Phi^2}.
\end{eqnarray}
Since these two types of interactions can interfere with each other, adding up their cross sections directly is not allowed. Considering the interference effect~\cite{Belanger:2008sj}, the $SN$ and $S^\dagger N$ scatter cross sections can be obtained as
\begin{eqnarray}
\sigma_{SN} = \frac{\mu_{SN}^2 f_{SN}^2}{\pi}, \quad \sigma_{S^\dagger N} = \frac{\mu_{S N}^2 f_{S^\dagger N}^2}{\pi},
\end{eqnarray}	
where
\begin{eqnarray}
f_{SN} = \frac{F_{S,N}}{2m_S} + F_{V,N}, \quad f_{S^\dagger N} = \frac{F_{S,N}}{2m_S} - F_{V,N},
\end{eqnarray}	
and $\mu_{SN}$ is the reduced mass of $S$ and $N$, defined by $\mu_{SN} \equiv m_S m_N/(m_S + m_N)$.
As we can see, $\sigma_{SN}$ and $\sigma_{S^\dagger N}$ are generally different from each other due to the isospin violation. As for the symmetric DM case, the average cross section of DM-nucleus point-like scattering can be written as
\begin{eqnarray}
\sigma_{DM-A} = \frac{\mu_{SN}^2}{2\pi} \Big\{ \big[ Z f_{Sp} + (A-Z)f_{Sn}  \big]^2 +  \big[ Z f_{S^\dagger p} + (A-Z)f_{S^\dagger n}  \big]^2 \Big\}.
\end{eqnarray}	
Here, we suppose the nucleus $A$ contains $Z$ protons and $(A-Z)$ neutrons. For the direct detection experiment, the detection material is dual-phase xenon, of which $Z$ equals 54. Considering that xenon has several isotopes $A_i$, the normalized-to-nucleon scattering cross section can be obtained as~\cite{Lao:2020inc}
\begin{eqnarray}
\sigma_N^Z = \sigma_{Sp} \frac{\sum_i \eta_i \mu_{SA_i}^2 \Big\{ \big[ Z + (A_i-Z)f_{Sn}/f_{Sp}  \big]^2 +  \big[ Z f_{S^\dagger p}/f_{Sp} + (A_i-Z)f_{S^\dagger n}/f_{Sp}  \big]^2 \Big\} }{2\sum_i \eta_i \mu_{SA_i}^2 A_i^2},
\end{eqnarray}	
where $\eta_i$ is the fractional abundance of isotope $A_i$ in nature. For xenon, $A_i = [128, 129, 130, 131, \\
132, 134, 136]$, $\eta_i = [1.9\%, 26\%, 4.1\%, 21\%, 27\%, 10\%, 8.9\%]$ for each corresponding isotope~\cite{Feng:2011vu}.

\begin{figure}[!h]
\centering
\subfigure[\label{fig3-1}]
{\includegraphics[width=0.48\textwidth]{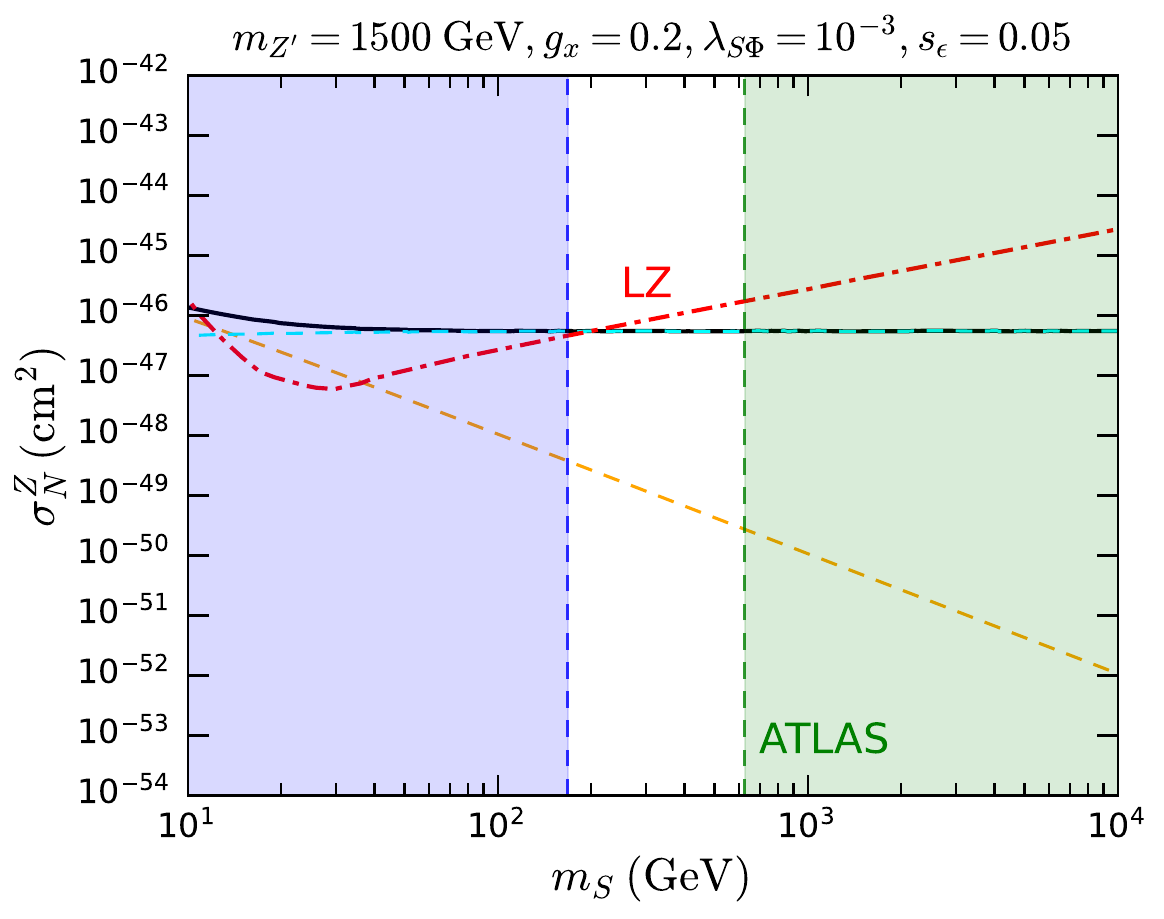}}
\hspace{.01\textwidth}
\subfigure[\label{fig3-2}]
{\includegraphics[width=0.48\textwidth]{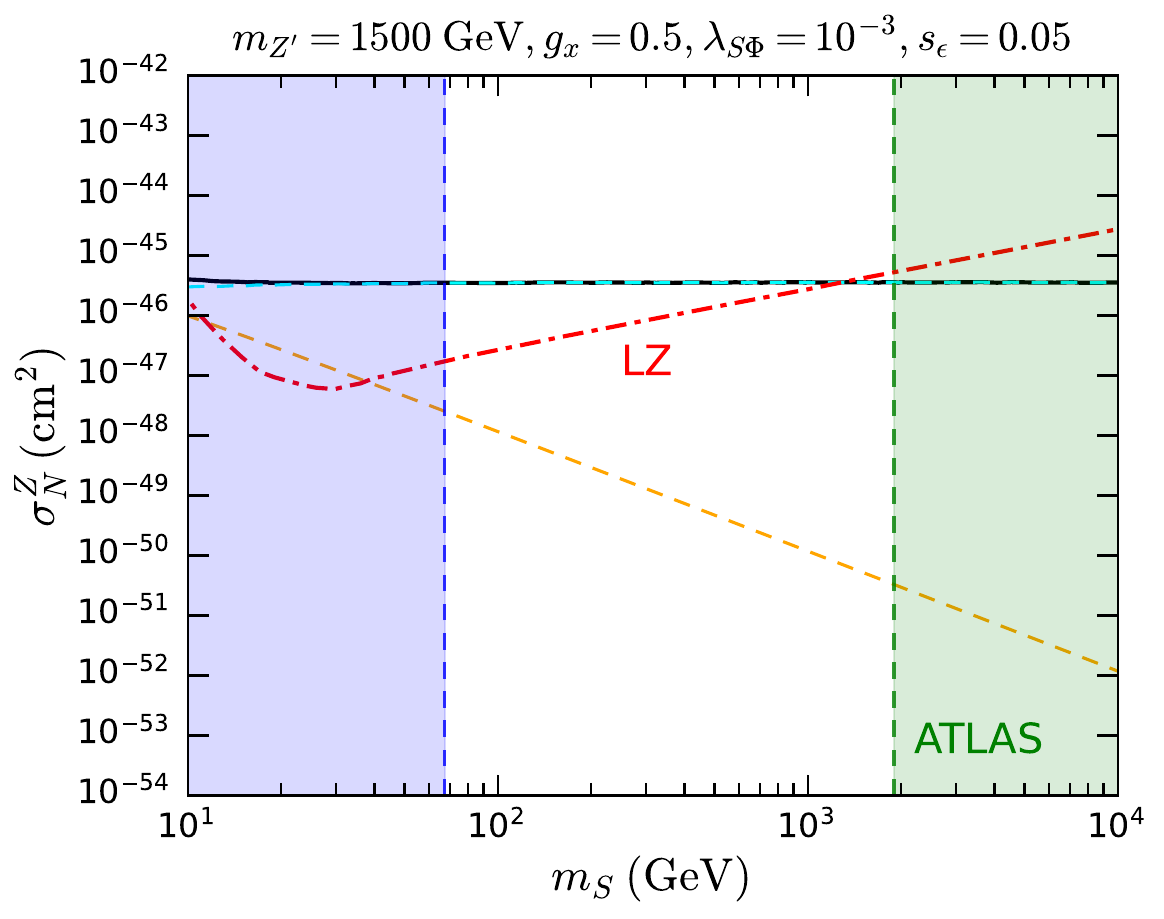}}
\hspace{.01\textwidth}
\subfigure[\label{fig3-3}]
{\includegraphics[width=0.48\textwidth]{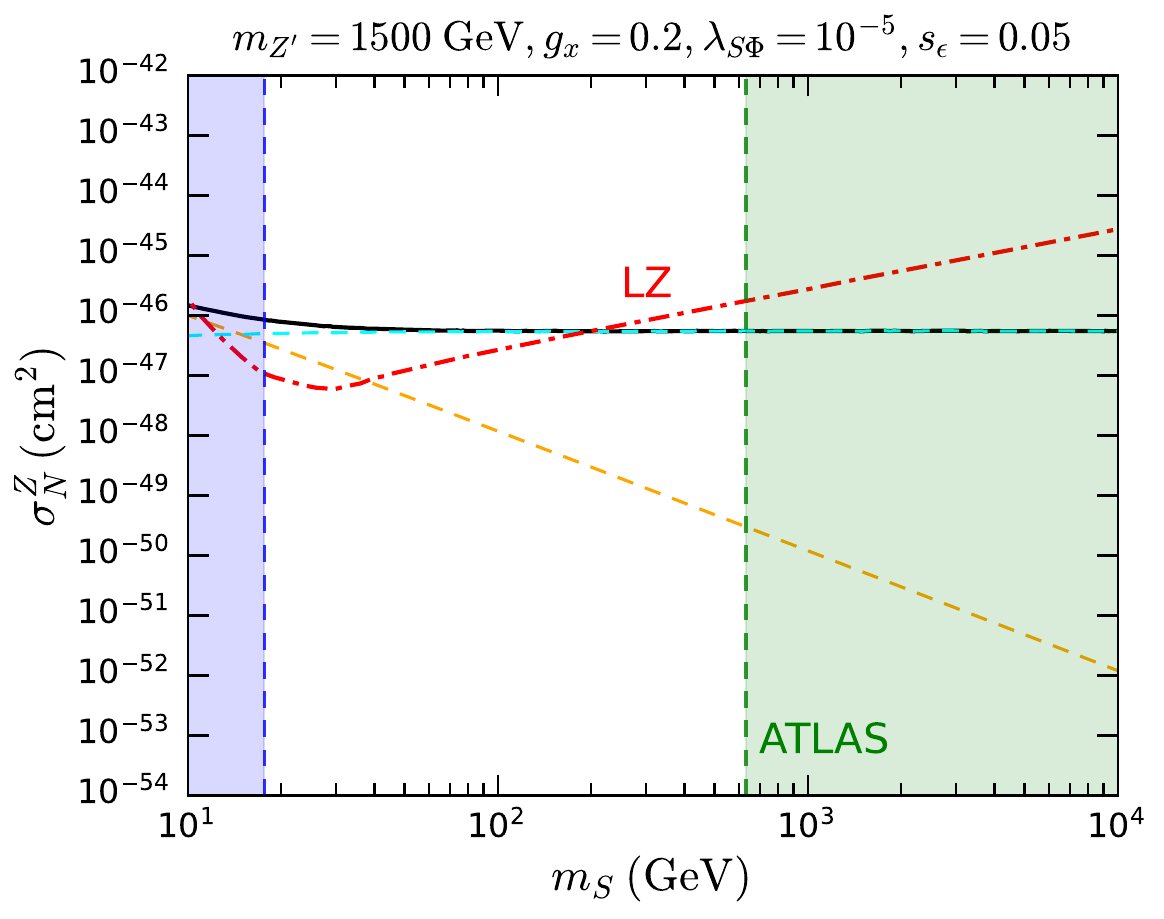}}
\subfigure[\label{fig3-4}]
{\includegraphics[width=0.48\textwidth]{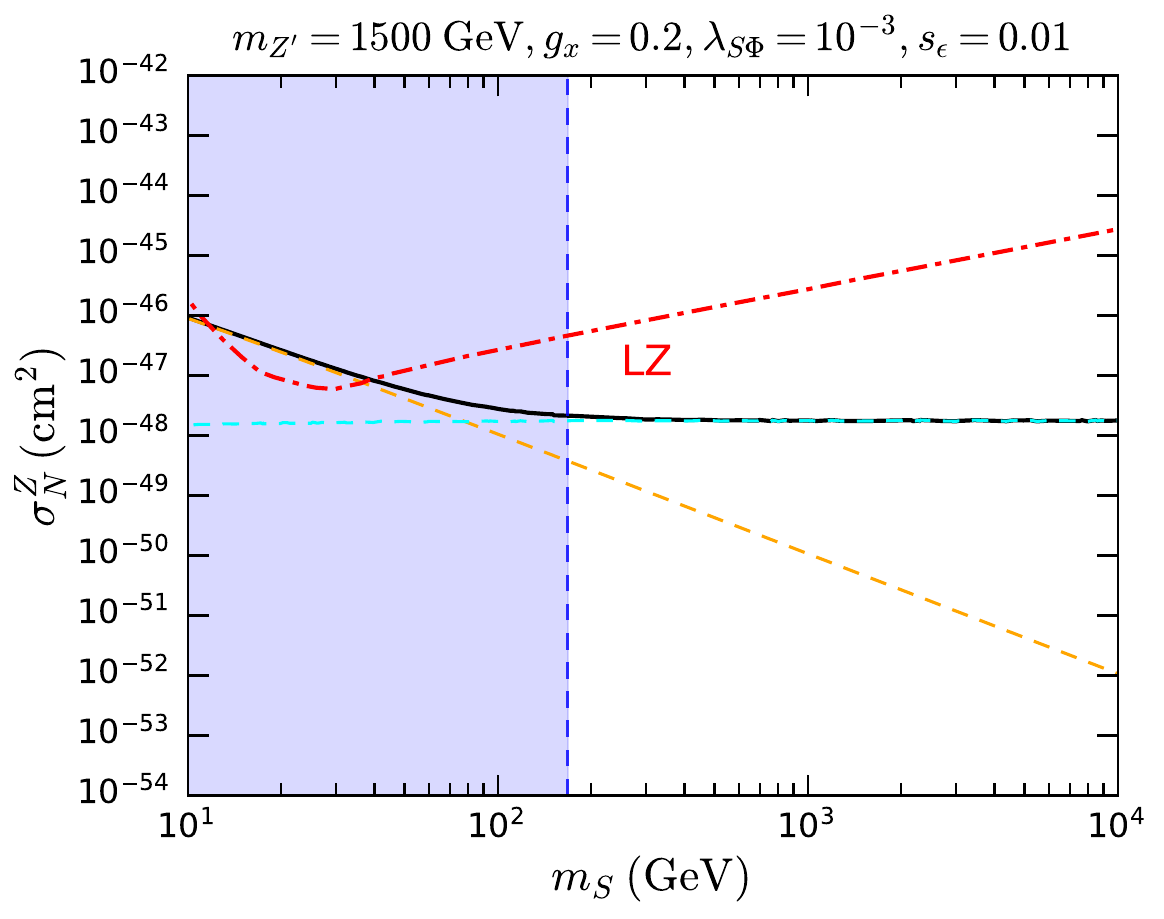}}
\caption{SI DM-nucleon scattering cross section $\sigma_N^Z$ as a function of $m_S$ for changed $g_x$ (b), $\lambda_{S\Phi}$ (c) and $s_\epsilon$ (d) compared to (a). The red line denotes the $90\%$ C.L. constraint from the LZ direct detection experiment. The blue and green regions are excluded by the DM stability and $Z'$ search experiment from ATLAS. The orange, cyan and black lines indicate the scattering cross section contributed from scalar interaction, vector interaction and the combination of them. }
\label{fig3}
\end{figure}

\begin{figure}[!h]
\centering
\subfigure[\label{fig4-1}]
{\includegraphics[width=0.48\textwidth]{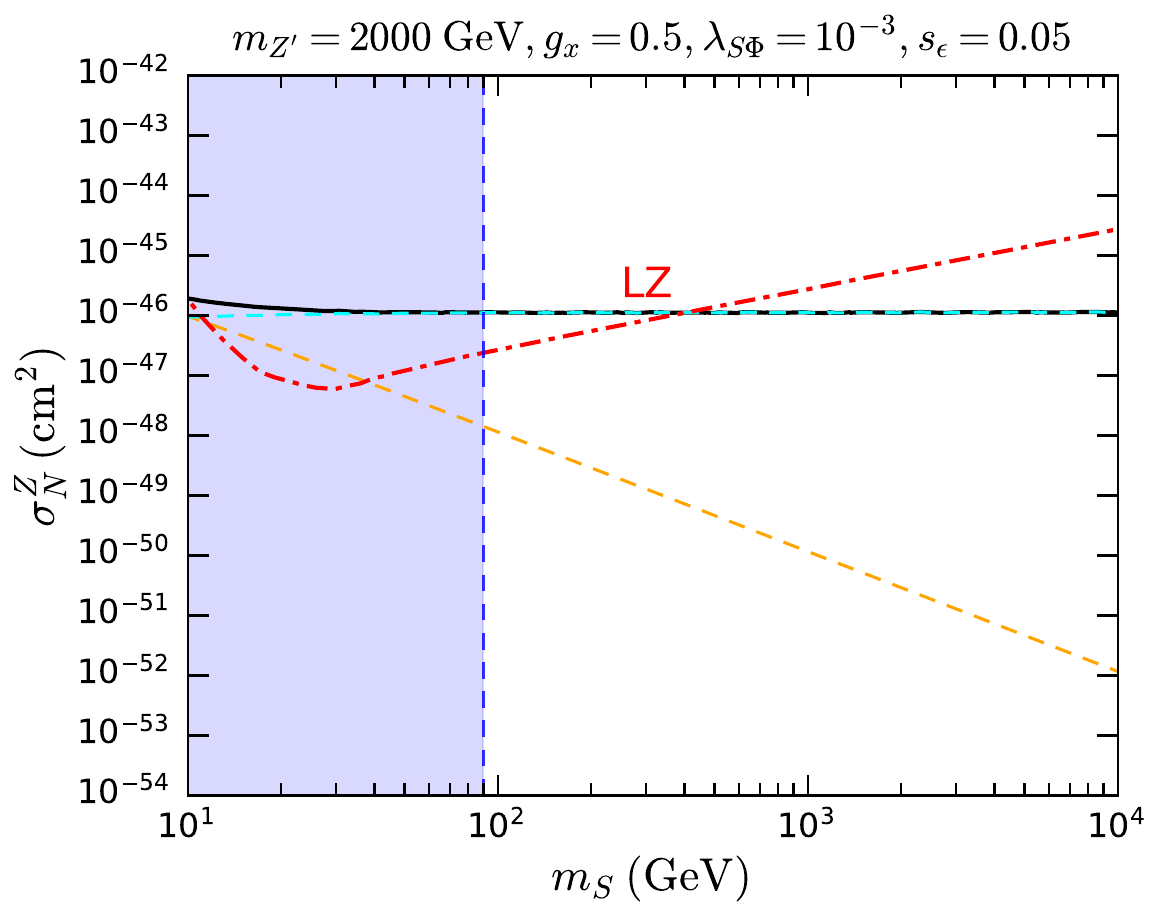}}
\hspace{.01\textwidth}
\subfigure[\label{fig4-2}]
{\includegraphics[width=0.48\textwidth]{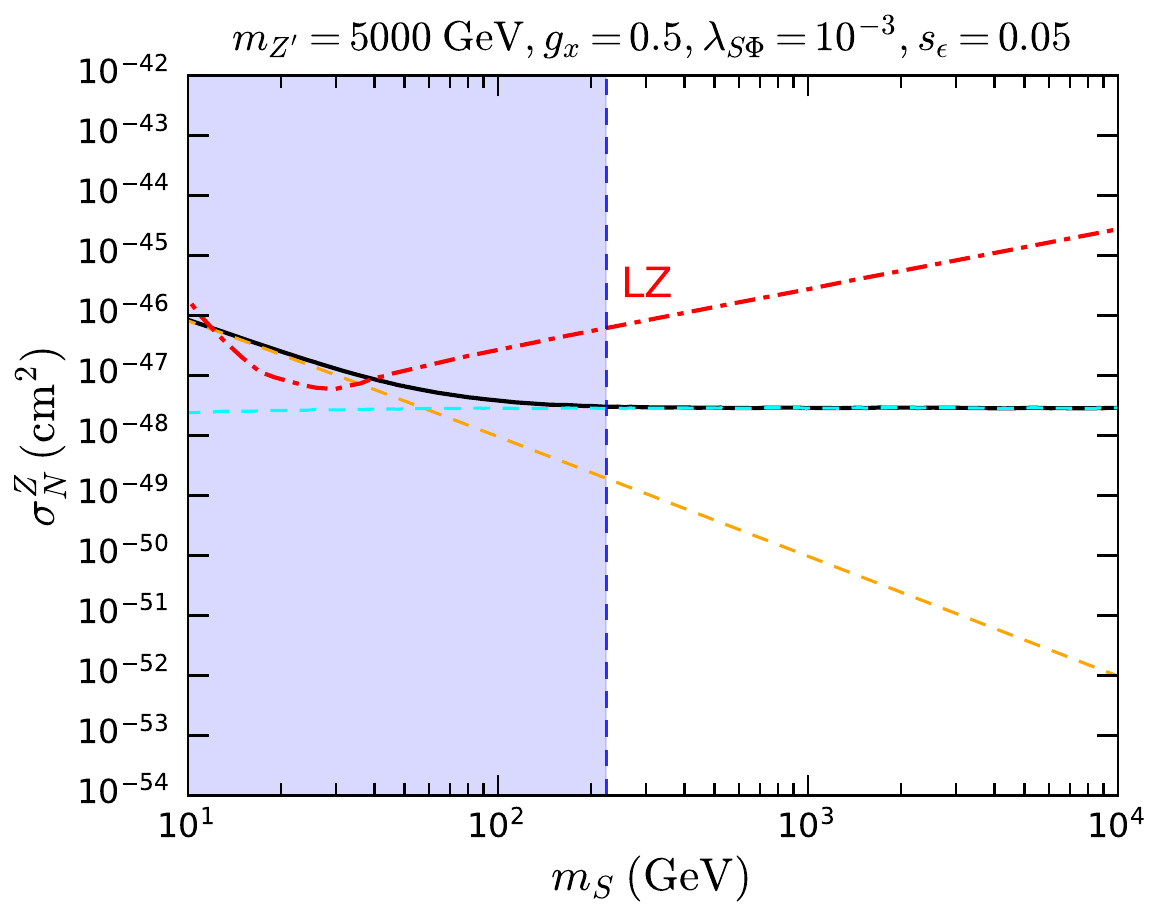}}
\caption{$\sigma_N^Z$ as a function of $m_S$ for changed $m_{Z'}$ to be $2000~\mathrm{GeV}$ (a) and $5000~\mathrm{GeV}$ (b). }
\label{fig4}
\end{figure}

In the subsequent discussion, we select certain benchmark scenarios outlined in Table~\ref{tab:direct_detection} to illustrate the SI DM-nucleon scattering cross section $\sigma_N^Z$. We maintain other parameters unchanged: $\lambda_S = \lambda_{HS} = \lambda_{H\Phi} = 10^{-3}$, $m_{h_2} = 1~\mathrm{TeV}$, and $n_\Phi = 10^{-3}$. The shaded blue area is excluded due to the stability of DM, while the green area is ruled out by ATLAS $Z'$ search experiment. Additionally, the direct detection limit from the LZ experiment at $90\%$ CL is depicted in the figure~\cite{LZ:2022ufs}. The black line denotes $\sigma_N^Z$, whereas the orange and cyan lines correspond to the scattering cross sections for scalar and vector interactions, respectively. It can be seen that at low/high $m_S$, the scalar/vector interaction dominates the DM-nucleon scattering process~\cite{Yu:2011by}. This trend is further elucidated in Figs.~\ref{fig3-4} and \ref{fig4-2}. 

In Fig.~\ref{fig3-1} with $m_{Z'} = 1500~\mathrm{GeV}$, $g_x = 0.2$, $\lambda_{S\Phi} = 10^{-3}$ and $s_\epsilon = 0.05$, the DM mass range between $12$ and $626~\mathrm{GeV}$ remains unexcluded. As $g_x$ increases in Fig.~\ref{fig3-2}, the constraints from DM stability and ATLAS weaken, but the DM-nucleon interaction strength intensifies. Fig.~\ref{fig3-3} demonstrates that relaxing the constraint from DM stability is achievable by reducing $\lambda_{S\Phi}$. The lower limit can be estimated as $5.5~\mathrm{GeV}$ by setting $\lambda_{S\Phi} = 0$. In Fig.~\ref{fig3-4} and Fig.~\ref{fig4} as $s_\epsilon$ decreases or $m_{Z'}$ increases, $Z'$ tends to decouple from SM particles, resulting in a reduction of $\sigma_N^Z$.

\begin{table}[htb]
\centering
\resizebox{0.48\textwidth}{!}{
\begin{tabular}{|c|c|c|c|c|c|}
\hline
&$m_{Z'}(\mathrm{GeV})$ & $g_x$ & $\lambda_{S\Phi}$ & $s_\epsilon$ & $m_{S}(\mathrm{GeV})$  \\ \hline
Fig.~\ref{fig3-1} & 1500 & 0.2 & $10^{-3}$ & $0.05$ & $12 \sim 626$    \\ \hline
Fig.~\ref{fig3-2} & 1500 & 0.5 & $10^{-3}$ & $0.05$ & $1280 \sim 1896$ \\ \hline
Fig.~\ref{fig3-3} & 1500 & 0.2 & $10^{-5}$ & $0.05$ &  $202 \sim 631$  \\ \hline
Fig.~\ref{fig3-4} & 1500 & 0.2 & $10^{-3}$ & $0.01$ &  $\gtrsim 168$  \\ \hline
Fig.~\ref{fig4-1} & 2000 & 0.5 & $10^{-3}$ & $0.05$ &  $\gtrsim 90$  \\ \hline
Fig.~\ref{fig4-2} & 5000 & 0.5 & $10^{-3}$ & $0.05$ &  $\gtrsim 224$  \\ \hline
\end{tabular}}
\caption{Surviving region of $m_{S}$ in each benchmark.}
\label{tab:direct_detection}
\end{table}

\subsection{DM annihilation }
The relic abundance of DM is determined by the annihilation cross section $\langle \sigma_{\mathrm{ann}} v \rangle_{\mathrm{FO}}$ at the freeze-out epoch. The primary annihilation channels include $ZZ$, $Z'Z'$, $W^+W^-$ and $h_i h_j$ ($i, j = 1,2$) in this model if kinematically allowed. For the reason that the couplings between SM fermions and Higgs are compressed by their masses, the annihilation channels into $f\bar{f}$ are strongly suppressed except for $t\bar{t}$. 
 
We compute the relic density and thermal-averaged cross section using \texttt{micrOMEGAs}~\cite{Belanger:2014vza}, a code capable of calculating DM properties within any given new physics model. The relic density must satisfy the Planck measurement $\Omega_{\mathrm{DM}}h^2 = 0.1200 \pm 0.0012$~\cite{Planck:2018vyg}. Parameter regions yielding a larger DM relic abundance than this limit are excluded, while those with smaller relic abundance are permissible, as other DM candidates or non-thermal production mechanisms may contribute to the total relic abundance. The annihilation cross section of DM during the freeze-out epoch is typically expected to be around $\mathcal{O}(10^{-26})~\mathrm{cm^3 s^{-1}}$, with a thermally averaged DM velocity of approximately $\mathcal{O}(10^{-1})$, to be consistent with the Planck result.

Currently, DM in the present universe is still undergoing annihilation, emitting gamma rays in the process. A joint analysis of MAGIC observations for 158 hours and Fermi Large Area Telescope (LAT) observations for 6 years has placed strong constraints on the DM annihilation cross section~\cite{MAGIC:2016xys}. Further studies indicate that DM particles in dwarf galaxies typically exhibit velocities on the order of $\mathcal{O}(10^{-5})$~\cite{2009ApJ...704.1274W}, suggesting that DM annihilation today can be well approximated as s-wave annihilation. In Section \ref{Parameter scan}, we utilize both the Planck measurement and the Fermi-MAGIC result to constrain our model parameters.

\section{Parameter scan}
\label{Parameter scan}
In this model, there are 10 free parameters, as delineated in Eq.~(\ref{free parameters}). We conduct a random scan within the following parameter ranges:
\begin{eqnarray}
&& 10\ \mathrm{GeV}<m_{h_2},\ m_S <10^{4}\ \mathrm{GeV},\quad 10^{2}\ \mathrm{GeV}<m_{Z'}<10^{4}\ \mathrm{GeV}, \nonumber\\
&& 10^{-2} < g_x < 1,\quad 10^{-3} < s_\varepsilon <0.9,\quad 10^{-3} < n_\Phi <1, \nonumber\\
&& 10^{-4}< \lambda_{S},\ \lambda_{HS},\ \lambda_{H\Phi},\ \lambda_{S\Phi} <1.
\end{eqnarray}
Moreover, we employ the following constraints to filter each parameter point:
\begin{itemize}
\item According to the \texttt{Lilith} calculation, p-value must exceed $0.05$. 
\item The SI DM-nucleon scattering cross section $\sigma_N^Z$ must not surpass the limit established by the LZ direct detection experiment.
\item The predicted relic density of DM, $\Omega_{\mathrm{DM}} h^2$, must fall within the Planck result.
\item The thermal-avearged annihilation cross section of DM in dwarf galaxies, $\langle \sigma_{\mathrm{ann}} v \rangle_{\mathrm{D}}$, should satisfy the Fermi-LAT constraint.
\end{itemize}

\begin{figure}[!ht]
\centering
\includegraphics[width=0.7\textwidth]{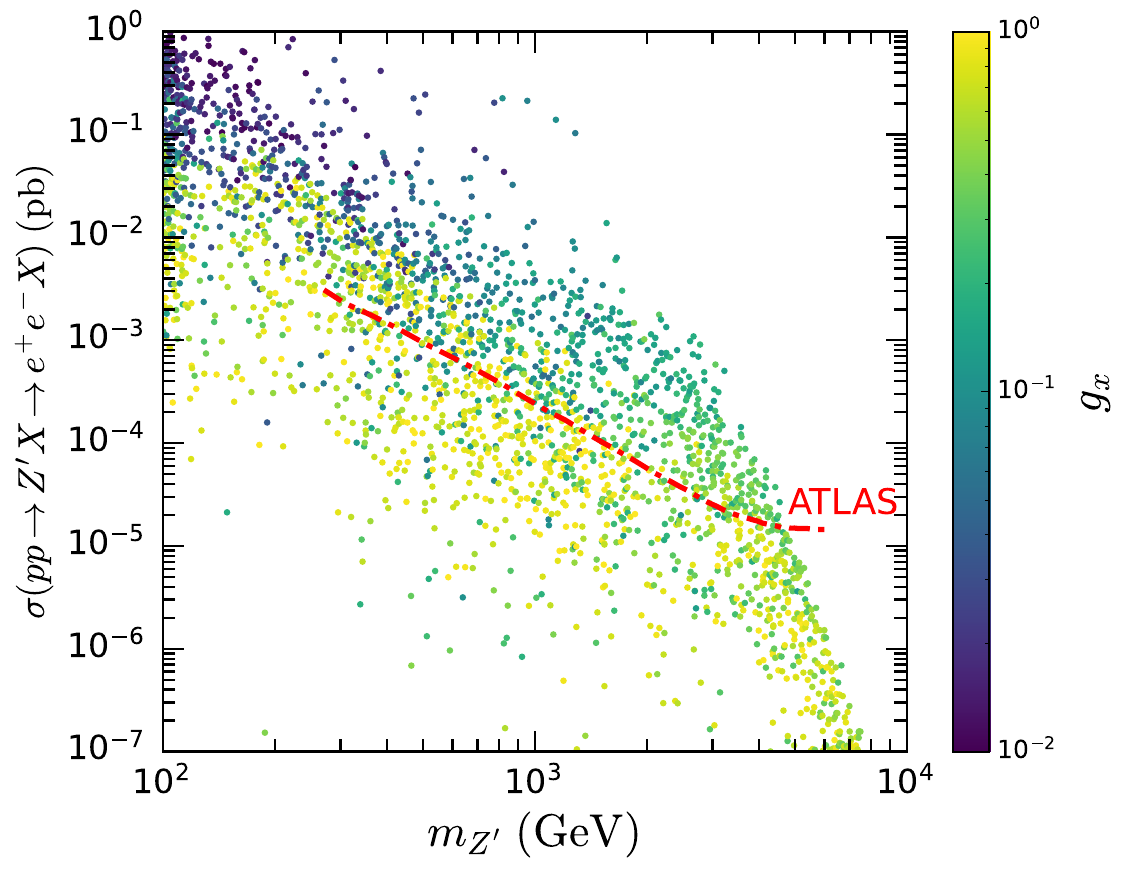}
\caption{ Selected parameter points projected onto the $\sigma(p p \to Z' X \to e^+ e^- X)$-$m_{Z'}$ plane, with a colorbar corresponding to $g_x$. The red line denotes the $95\%$ CL upper limit set by ATLAS. }
\label{fig6}
\end{figure}
In Fig.~\ref{fig6}, we project the parameter points satisfying the above constraints onto the $\sigma(p p \to Z' X \to e^+ e^- X)$-$m_{Z'}$ plane, with a colorbar corresponding to $g_x$ displayed on the right. Additionally, we include the $95\%$ CL upper limit set by ATLAS on the graph. It can be seen that for $m_{Z'} \gtrsim 4750~\mathrm{GeV}$, all parameter points can evade detection in the $Z'$ search experiment. Conversely, for $m_{Z'} \lesssim 4750~\mathrm{GeV}$, smaller values of $g_x$ are disfavored. This trend emerges because, as $g_x$ increases, the interaction between $Z'$ and SM particles weakens. For the majority of surviving parameter points, $g_x$ ranges from $0.1$ to $1$. Utilizing interpolation, we discard parameter points that exceed the upper limit, ensuring that the remaining points satisfy all considered constraints. Finally, the last phenomenological requirement is
\begin{itemize}
\item The cross section $\sigma(p p \to Z' X \to e^+ e^- X)$ for producing $Z'$ at the LHC, which then decays to $e^+ e^-$ final state, falls within the ATLAS limit.
\end{itemize}
These criteria are essential for ensuring that the model remains consistent with both theoretical expectations and experimental observations.

\begin{figure}[!h]
\centering
\subfigure[\label{fig7-1}]
{\includegraphics[width=0.48\textwidth]{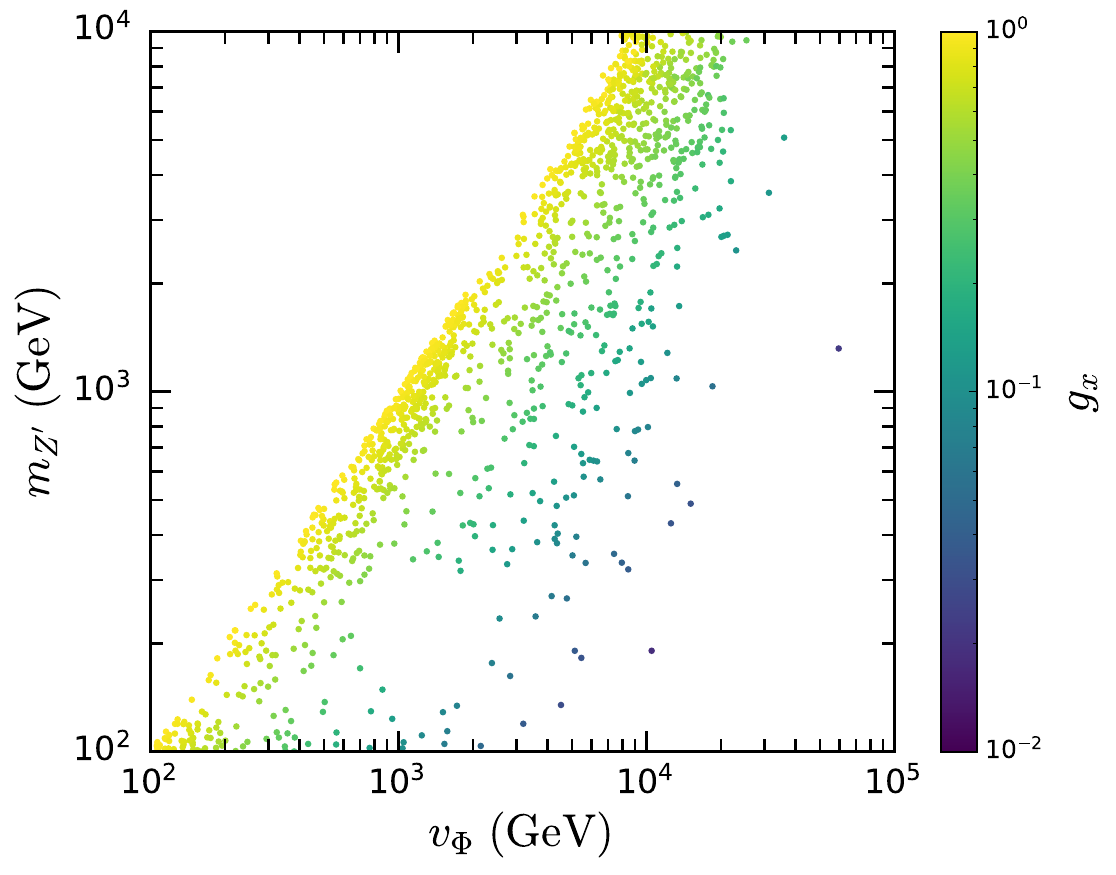}}
\hspace{.01\textwidth}
\subfigure[\label{fig7-2}]
{\includegraphics[width=0.48\textwidth]{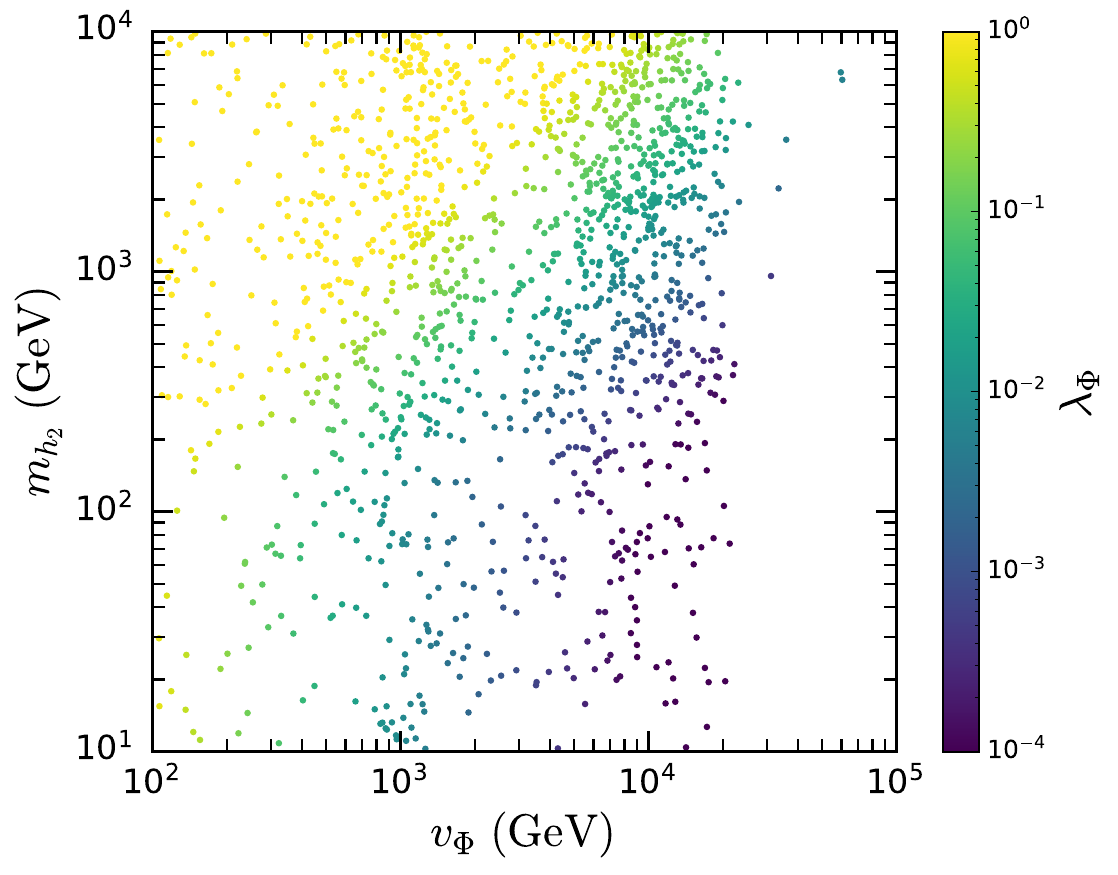}}
\hspace{.01\textwidth}
\subfigure[\label{fig7-3}]
{\includegraphics[width=0.48\textwidth]{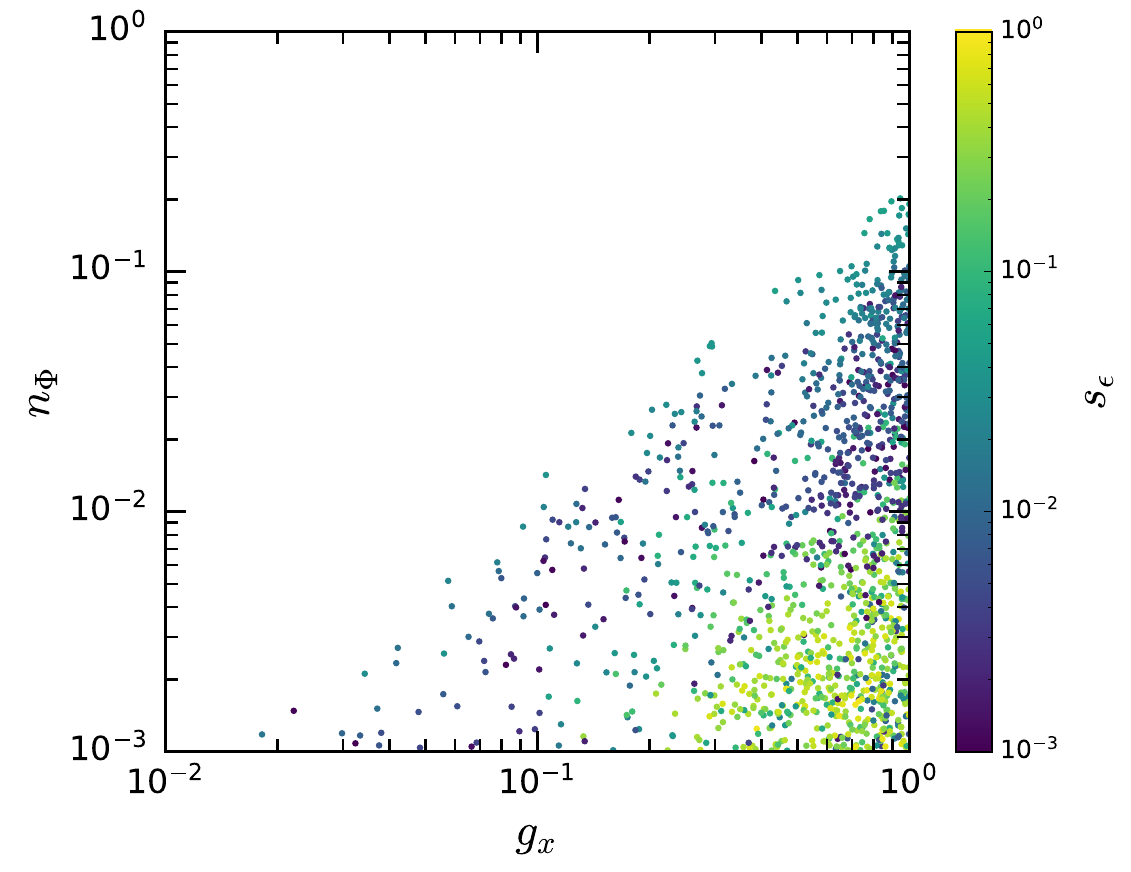}}
\caption{Surviving parameter points projected onto the the $v_\Phi$-$m_{Z'}$ (a), $v_\Phi$-$m_{h_2}$ (b) and $g_x$-$n_{\Phi}$ (c) planes, with colorbars corresponding to $g_x$, $\lambda_\Phi$ and $s_\epsilon$, respectively. }
\label{fig7}
\end{figure}
Then we project the surviving parameter points onto the $v_\Phi$-$m_{Z'}$, $v_\Phi$-$m_{h_2}$ and $g_x$-$n_{\Phi}$ planes in Figs.~\ref{fig7-1}, \ref{fig7-2} and \ref{fig7-3}, with colorbars corresponding to $g_x$, $\lambda_\Phi$ and $s_\epsilon$, respectively. Figs.~\ref{fig7-1} and \ref{fig7-2} suggest that $m_{Z'}$ and $m_{h_2}$ have an approximate linear relationship with $v_\Phi$. Particularly, this relationship appears more pronounced in Fig.~\ref{fig7-1}, implying $m_{Z'} \simeq g_x v_\Phi$. This is consistent with the case when $t'_\epsilon \to 0$ in Eq.~(\ref{mass:mZp}). In contrast, when $m_{h_2} \lesssim 200~\mathrm{GeV}$ in Fig.~\ref{fig7-2}, such linearity tends to break down. According to Eq.~(\ref{mass:higgs}), when the masses of $h_1$ and $h_2$ become close, this approximate relationship no longer holds. Fig.~\ref{fig7-3} illustrates the relation among the parameters in the gauge sector. It shows that $n_\Phi$ is proportional to $g_x$ and negatively correlated to $s_\epsilon$. More precisely, $g_x$ ranges from approximately $0.02$ to $1$ while $n_\Phi$ in $10^{-3} \sim 0.2$ and $s_\epsilon$ in $10^{-3} \sim 0.9$. This surviving parameter space in the gauge sector can be further examined by future $Z'$ searches and direct detection experiments.

\begin{figure}[!h]
\centering
\subfigure[\label{fig8-1}]
{\includegraphics[width=0.48\textwidth]{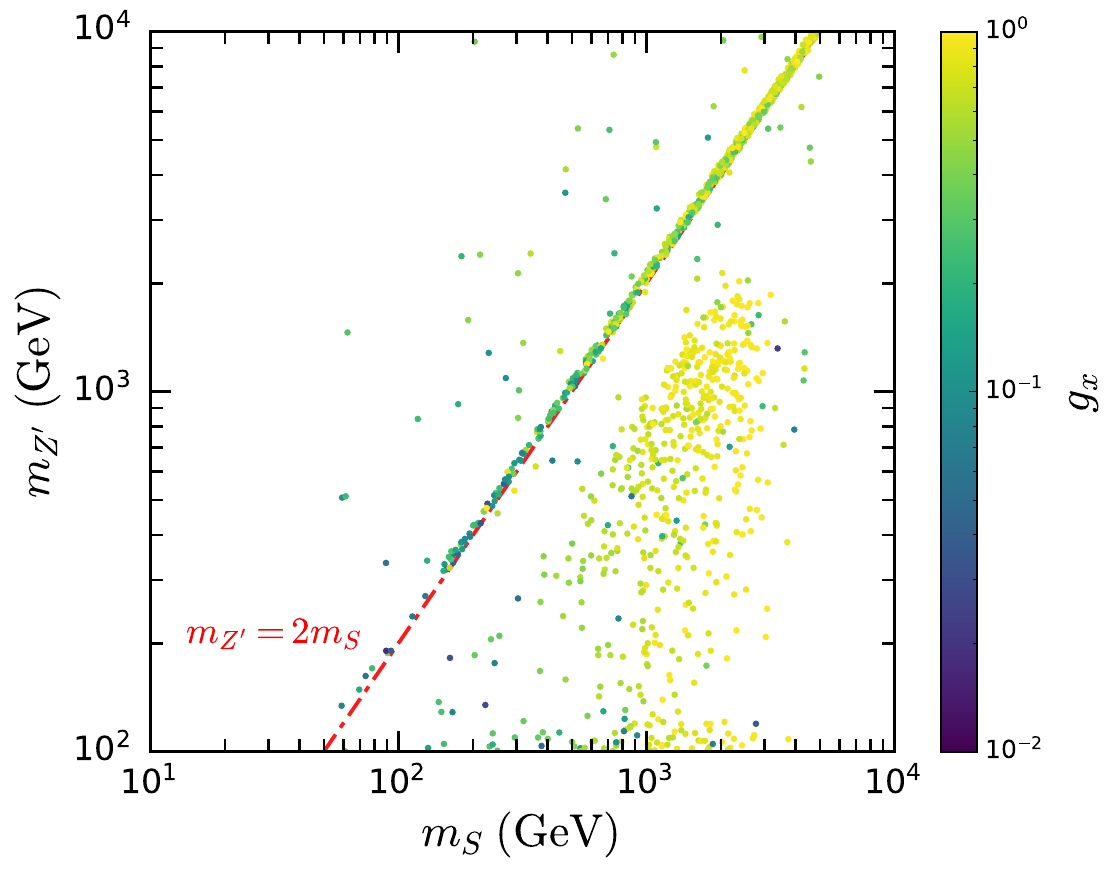}}
\hspace{.01\textwidth}
\subfigure[\label{fig8-2}]
{\includegraphics[width=0.48\textwidth]{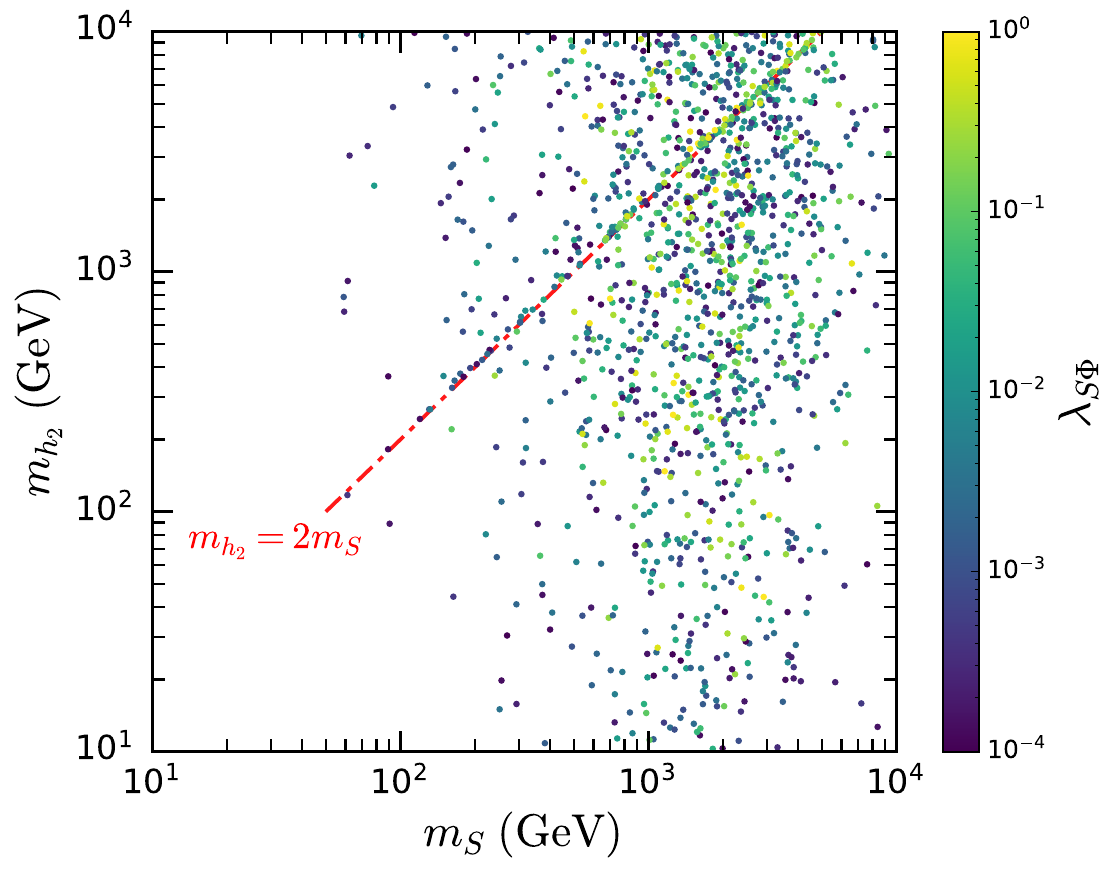}}
\caption{Surviving parameter points projected onto the the $m_S$-$m_{Z'}$ (a) and $m_{S}$-$m_{h_2}$ (b) planes, with colorbars corresponding to $g_x$ and $\lambda_{S\Phi}$, respectively. The red lines in (a) and (b) correspond to the resonant annihilation of DM via mediators $Z'$ and $h_2$, respectively.  }
\label{fig8}
\end{figure}
Figs.~\ref{fig8-1} and \ref{fig8-2} illustrate the resonant annihilation of DM in the the $m_S$-$m_{Z'}$ and $m_{S}$-$m_{h_2}$ planes, with colors denoting $g_x$ and $\lambda_{S\Phi}$. In Fig.~\ref{fig8-1},  a cluster of parameter points concentrates along the line where $m_{Z'} = 2m_S$, representing the resonance region where two DM particles annihilate into one $Z'$ particle. By comparison, in Fig.~\ref{fig8-2}, only a few points concentrate along the line where $m_{h_2} = 2m_S$, indicating that the resonant annihilation mediated by $h_2$ is much weaker than that mediated by $Z'$ for most parameter points.

\begin{figure}[!h]
\centering
\subfigure[\label{fig9-1}]
{\includegraphics[width=0.48\textwidth]{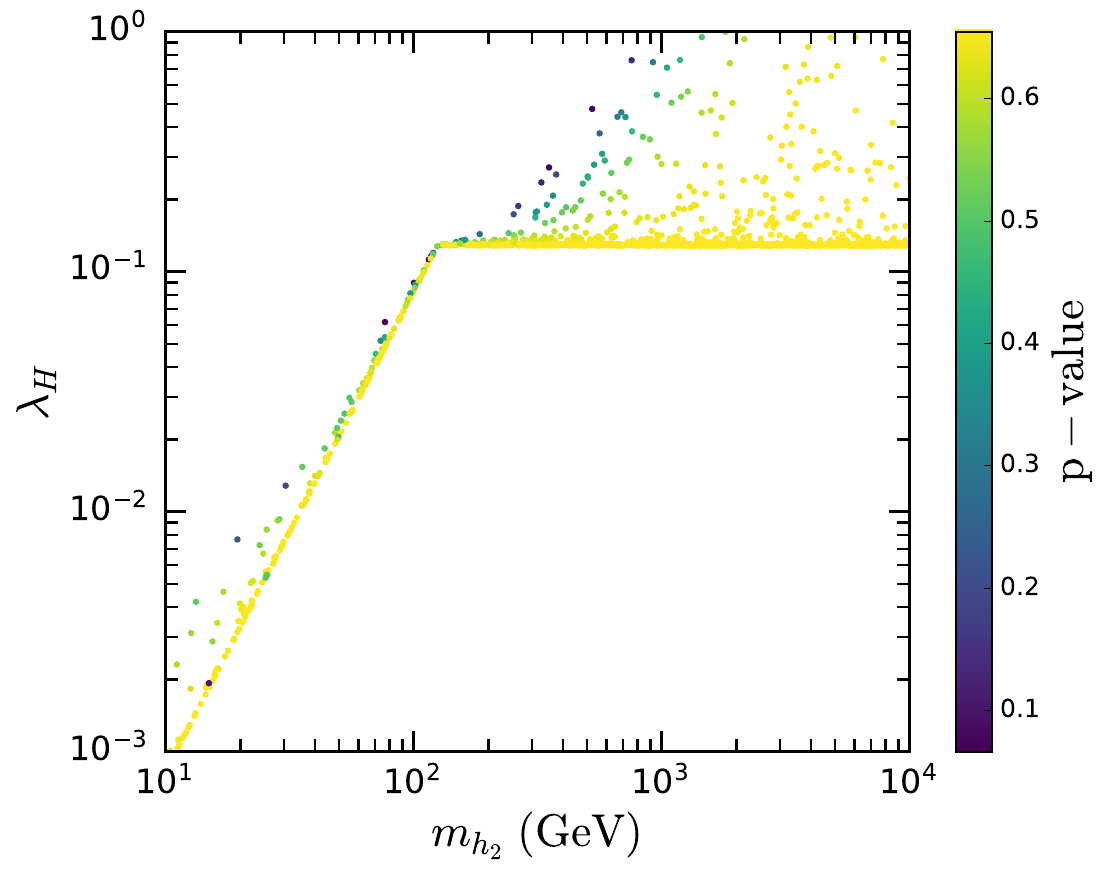}}
\hspace{.01\textwidth}
\subfigure[\label{fig9-2}]
{\includegraphics[width=0.48\textwidth]{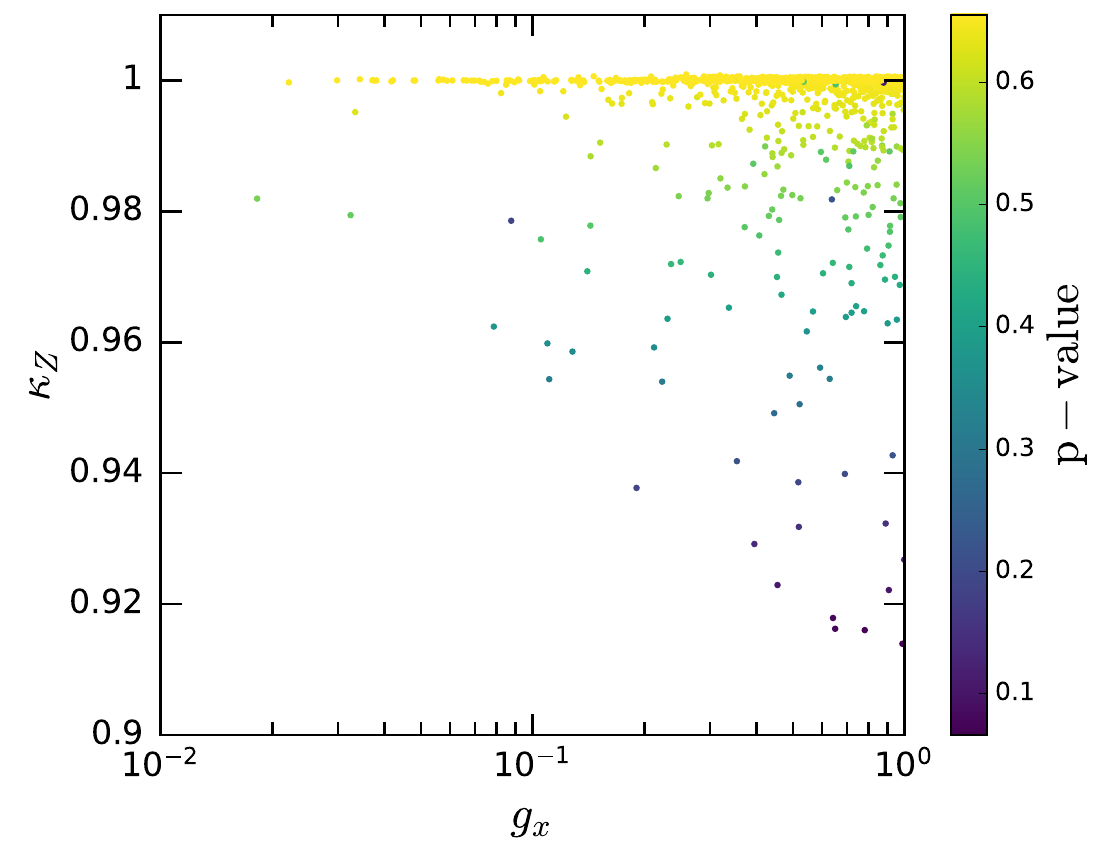}}
\caption{Surviving parameter points projected onto the $m_{h_2}$-$\lambda_{H}$ and $g_x$-$\kappa_{Z}$ planes, with colorbars corresponding to the Lilith p-value of each point.  }
\label{fig9}
\end{figure}
Fig.~\ref{fig9} depicts the Lilith p-values of the surviving parameter points projected onto the $m_{h_2}$-$\lambda_{H}$ and $g_x$-$\kappa_{Z}$ planes. In Fig.~\ref{fig9-1} when $
m_{h_2} \gtrsim m_{h_1}$, $\lambda_H$ tends to converge toward $\lambda_{SM} = m_{h_1}^2/(2v_H^2) \simeq 0.13$, which corresponds to the quartic Higgs coupling in the SM. Therefore, along this linear trend lies the region with the largest p-value.  Conversely, when $m_{h_2} \lesssim m_{h_1}$, $\lambda_H$ deviates significantly from $\lambda_{SM}$ and converges along the line described by Eq.~(\ref{mass:higgs}). In Fig.~\ref{fig9-2},  the majority of parameter points converge around $\kappa_Z = 1$, consistent with the SM prediction. Furthermore, most points tend to cluster within the region where $g_x$ ranges from $0.1$ to $1$. This trend arises because smaller values of $g_x$ are more likely to be excluded by $Z'$ search experiments, aligning with the earlier discussion.

\begin{figure}[!h]
\centering
\subfigure[\label{fig10-1}]
{\includegraphics[width=0.48\textwidth]{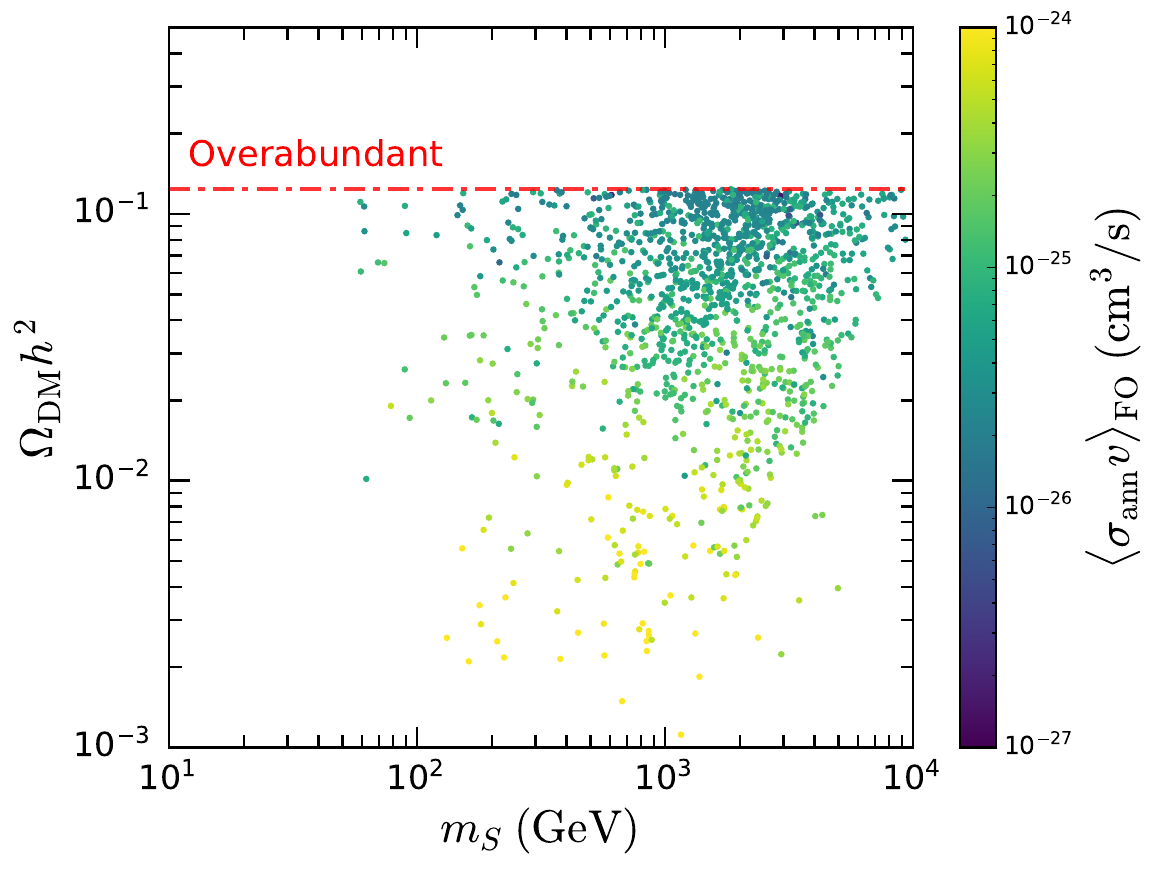}}
\hspace{.01\textwidth}
\subfigure[\label{fig10-2}]
{\includegraphics[width=0.48\textwidth]{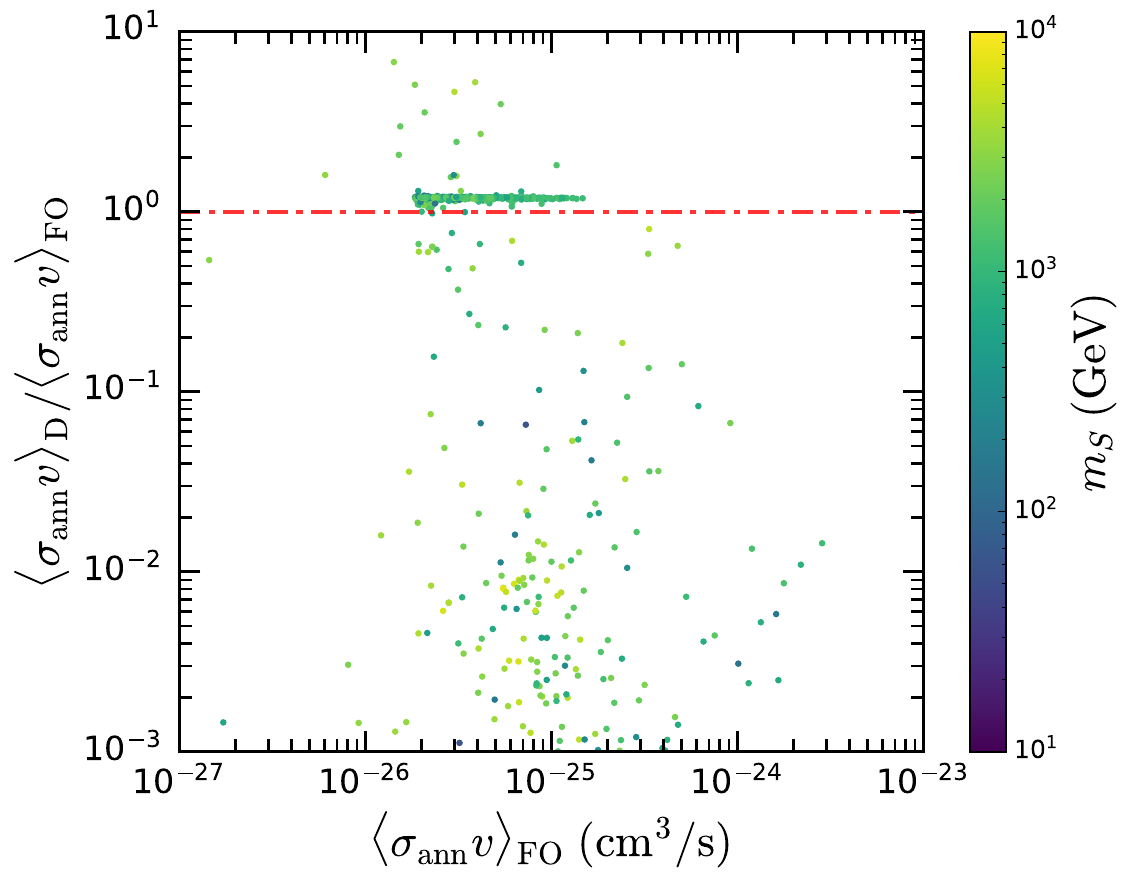}}
\caption{Surviving parameter points projected onto the  $m_{S}$-$\Omega_{\mathrm{DM}}h^2$ and  $\langle \sigma_{\mathrm{ann}} v \rangle_{\mathrm{FO}}$-$\langle \sigma_{\mathrm{ann}} v \rangle_{\mathrm{FO}}/\langle \sigma_{\mathrm{ann}} v \rangle_{\mathrm{D}}$ planes, with colorbars corresponding to $\langle \sigma_{\mathrm{ann}} v \rangle_{\mathrm{FO}}$ and $m_S$ respectively. The red line in (a) denotes the Planck result while the one in (b) means that the parameter points around it are s-waves dominated.}
\label{fig10}
\end{figure}
Moreover, we present the surviving parameter points in the $m_{S}$-$\Omega_{\mathrm{DM}}h^2$ plane in Fig.~\ref{fig10-1}, with a colorbar indicating the thermal-avearged annihilation cross section  $\langle \sigma_{\mathrm{ann}} v \rangle_{\mathrm{FO}}$ at the freeze-out epoch. The red line denotes the Planck observation, with the area above it representing the over-production of DM. In Fig.\ref{fig10-2}, we illustrate the relation between $\langle \sigma_{\mathrm{ann}} v \rangle_{\mathrm{FO}}$ and $\langle \sigma_{\mathrm{ann}} v \rangle_{\mathrm{D}}$. The red line denotes the case where $\langle \sigma_{\mathrm{ann}} v \rangle_{\mathrm{FO}} = \langle \sigma_{\mathrm{ann}} v \rangle_{\mathrm{D}}$. Most of the parameter points cluster around the vicinity of the red line, demonstrating that they are dominated by s-waves. The range of $\langle \sigma_{\mathrm{ann}} v \rangle_{\mathrm{FO}}$ for these points is approximately $10^{-26}\mathrm{cm^3/s}$ to $10^{-25}\mathrm{cm^3/s}$. However, there are also a number of parameter points far away from the red line, indicating significant differences between their $\langle \sigma_{\mathrm{ann}} v \rangle_{\mathrm{D}}$ and $\langle \sigma_{\mathrm{ann}} v \rangle_{\mathrm{FO}}$. This discrepancy arises because these points lie within the DM resonant annihilation region, resulting in a strong dependence on their velocity. Consequently, $\langle \sigma_{\mathrm{ann}} v \rangle_{\mathrm{FO}}$ for these points is no longer dominated by s-waves.

\section{Conclusions}
\label{conclusions}
In this work, we propose a scalar dark matter model with a hidden gauge symmetry $\UoneX$ and two complex scalars $\Phi$ and $S$. Two types of mixings between $\UoneX$ and $\UoneY$ are introduced to simultaneously generate dark matter relic abundance and satisfy the direct detection constraint. As the new gauge boson $Z'$ interacts with SM particles, our model is constrained from colliders as well. We discuss the constraint from electroweak precision measurements in the effective Lagrangian framework. Considering the significant difference between CDF measurement and other experiments, we fit our model parameters using two sets of $S$, $T$ and $U$ parameters for comparison. We present the fitting results for several benchmarks. We also discuss the constraint from the search for $Z'$ at the LHC. Our analysis suggests that large $m_S$ and small $s_\epsilon$ are favored. Especially, We present several surviving parameter spaces to be examined in future experiments. Additionally, the interaction within the Higgs sector may change in this model. Therefore, we select parameter points in the random scan conducted in Sec.\ref{Parameter scan}, to satisfy the constraints from the ATLAS and CMS Run 2 Higgs results. Considering the different couplings of $Z$ and $Z'$ to protons and neutrons, the isospin violation effect is considered in calculating the DM-nucleon scattering cross sections. Our findings show a wide range of the parameter space remains unexplored by current experiments. We present several parameter spaces by varying $s_\epsilon$, $m_{Z'}$, $g_x$, and $\lambda_{S\Phi}$. These parameter spaces will be further tested by future direct detection experiments.

By considering all the above constraints, and the Planck and Fermi-LAT observations as well, we perform a random scan to investigate the parameter space. We observe that DM annihilation mediated by $Z'$ is dominated compared with other mediators. The resonance effect renders annihilation sharply velocity-dependent in the resonance region, where $\langle \sigma_{\mathrm{ann}} v \rangle_{\mathrm{FO}} \neq \langle \sigma_{\mathrm{ann}} v \rangle_{\mathrm{D}}$. In the standard scenario where $\langle \sigma_{\mathrm{ann}} v \rangle_{\mathrm{FO}} \simeq \langle \sigma_{\mathrm{ann}} v \rangle_{\mathrm{D}}$, our scan results indicate that the thermal-averaged cross section at the freeze-out period $\langle \sigma_{\mathrm{ann}} v \rangle_{\mathrm{FO}} $ lies between $10^{-26}\mathrm{cm^3/s}$ and $ 10^{-25}\mathrm{cm^3/s}$. For the majority of surviving parameter points, $g_x$ ranges from $0.1$ to $1$. Additionally, we find that the regions where $m_{Z'} \gtrsim 4750~\mathrm{GeV}$ and $m_{Z'} \lesssim 4750~\mathrm{GeV}$ for $g_x$ close to $1$ remain viable and can be tested by future experiments.

\begin{acknowledgments}
CFCai is supported by the National Natural Science Foundation of China (NSFC) under Grant No. 11905300.
HHZhang is supported by NSFC under Grant No. 12275367.
YPZeng is supported by program for scientific research start-up funds of Guangdong Ocean University.
This work is also supported by the Fundamental Research Funds for the Central Universities, and the Sun Yat-sen University Science Foundation.

\end{acknowledgments}

\providecommand{\href}[2]{#2}\begingroup\raggedright

\end{document}